\documentclass[12pt]{article}
\usepackage{graphicx}
\usepackage{amsfonts}
\usepackage{amssymb,amsmath}
\usepackage{latexsym}
\usepackage{color}
\usepackage{cite}
\usepackage{bm}
\usepackage{ulem}
\input{colordvi.tex}

\renewcommand{\thefootnote}{\#\arabic{footnote}}

\begin{document}

\renewcommand{\thepage}{\arabic{page}}
\setcounter{page}{1}
\renewcommand{\thefootnote}{\#\arabic{footnote}}

\begin{titlepage}

\begin{center}

\vskip .5in

{\Large \bf 
Elucidating Dark Energy \\ with  Future 21 cm Observations
at the Epoch of Reionization
}

\vskip .45in

{\large
Kazunori~Kohri$\,^{1,2}$,
Yoshihiko~Oyama$\,^3$, 
Toyokazu~Sekiguchi$\,^4$ \\ and
Tomo~Takahashi$\,^5$
}

\vskip .45in

{\it
$^1$The Graduate University for Advanced Studies (Sokendai), \\ 1-1 Oho, Tsukuba 305-0801, Japan \vspace{2mm}\\
$^2$Institute of Particle and Nuclear Studies, KEK, 1-1 Oho, Tsukuba 305-0801, Japan \vspace{2mm}\\
$^3$Institute for Cosmic Ray Research, The University of Tokyo, \\5-1-5 Kashiwanoha, Kashiwa, Chiba 277-8582, Japan  \vspace{2mm}\\
$^4$Institute for Basic Science, Center for Theoretical Physics of the Universe, \\ Daejeon 34051, South Korea  \vspace{2mm}\\
$^5$Department of Physics, Saga University, Saga 840-8502, Japan
}

\end{center}

\vskip .4in

\begin{abstract}
  We investigate how precisely we can determine the nature of dark
  energy such as the equation of state (EoS) and its time dependence by
  using future observations of 21 cm fluctuations 
at the epoch of reionization ($6.8\lesssim z\lesssim10$)
such as Square
  Kilometre Array (SKA) and Omniscope in combination with those from
  cosmic microwave background, baryon acoustic oscillation, type Ia
  supernovae and direct measurement of the Hubble constant.  
We consider several parametrizations for the EoS 
and find that   future 21 cm observations will be powerful in constraining models of dark energy,
  especially when its EoS varies at high redshifts.
\end{abstract}
\end{titlepage}

\setcounter{footnote}{0}

%%%%%%%%%%%%%%%%%%%%%%%%%%%%%%%%%%%%%%%%%%%
%%%%%%%%%%%%%%%%%%%%%%%%%%%%%%%%%%%%%%%%%%%
\section{Introduction}

The origin of the present acceleration of the Universe has been one of
the biggest mysteries in modern cosmology.  To explain this
acceleration, one can assume a mysterious component called dark energy
\cite{DE_review} or, also resort to the modification of gravity
\cite{MG_review}.  Recently there have been attempts to describe these
models using the effective field theory
\cite{Gubitosi:2012hu,Gleyzes:2013ooa,Bloomfield:2012ff}.
Another  possible explanation  such as late-time quantum
backreaction from inflationary fluctuations has also been discussed
\cite{Ringeval:2010hf,Glavan:2014uga,Aoki:2014dqa,Glavan:2015cut}.

A lot of efforts have also been made to elucidate the nature of dark
energy by using cosmological observations.  When one investigates the
nature of dark energy using observational data, a phenomenological
approach is often taken, in which the properties of dark energy,
particularly, its equation of state (EoS) is parametrized in a general
way.  Since the EoS depends on time in most models of dark energy, its
time evolution is usually accommodated when parametrizing it and
constraints on such EoS parameters have been analyzed by using actual
cosmological data (see \cite{Ade:2015rim} for the constraint from the
measurement of cosmic microwave background (CMB) of Planck in
combination with baryon acoustic oscillation (BAO), type Ia supernovae
(SNe) and $H_0$ measurements for some parametrization, and
\cite{Qing-Guo:2016ykt} for a recent analysis).  Although
observational constraints on dark energy parameters are becoming more
and more severe, we are still far from pinning down the model which
describes the present-day cosmic acceleration.  However because the
accuracy of cosmological observations will be much improved in future,
the nature of dark energy can be probed more accurately. Therefore it
would be worth investigating how precisely we can obtain the
information on dark energy parameters in future cosmological
observations\footnote{
A lot of researches have been done in this direction, by adopting
future observations of CMB, large scale structure and so on,
see~\cite{Hollenstein:2009ph,Abdalla:2009wr,Debono:2009bd,Santos:2013gqa,Takeuchi:2014iza,Leung:2016xli} for relatively recent works along this line.
}.

As such, we in this paper focus on observations of fluctuations of 21
cm line of neutral hydrogen.  Observations of 21 cm line can probe
different redshift epochs compared with  other methods. Therefore
we can obtain unique information which cannot be acquired by  other
observations.  In addition, the next generation of the 21 cm survey,
Square Kilometre Array (SKA) \cite{ska}, will be in operation in
2020s. Hence it is timely to study expected constraints on dark energy
by using 21 cm experiments.  There have been several works on dark
energy using future observational data of 21 cm intensity mapping
\cite{Bull:2014rha,Xu:2014bya,Chen:2015vdi,Hossain:2016nlb}, HI galaxy survey
\cite{Zhao:2015wqa} and 21 cm fluctuations
\cite{Nusser:2004ps,Wyithe:2007rq,Pritchard:2008wy}.  In this paper,
we use the 21 cm fluctuations to derive projected constraints on
various dark energy models from SKA in combination with future CMB
experiments such as COrE+ \cite{core+} and future observations of SNe,
BAO and a direct measurement of Hubble constant $H_0$, assuming several
types of dark energy parametrizations. As mentioned above, because the
21 cm observations can probe the different epochs of the dark energy
evolution compared to other observations, it should be
complementary to  those of others. In addition to SKA, we also
consider the next-next generation of the 21 cm experiment such as
Omniscope \cite{Tegmark:2009kv}, which is expected to give an
unprecedented accuracy. By making an analysis adopting the above
mentioned observations, we discuss how the 21 cm observation can probe
the nature of dark energy.

The organization of this paper is as follows. In the next section, we
summarize how to parametrize dark energy EoS, which is
used in this paper. Then in Section \ref{sec:current_const}, we
present  current constraints on dark energy parameters introduced
in Section~\ref{sec:DE_param}. Then in Section~\ref{sec:future_const},
we investigate expected constraints on dark energy from future
observations of 21 cm line in combination with other observations such
as CMB, BAO, SNeIa and $H_0$.  The final section is devoted to the conclusion of this paper.

%%%%%%%%%%%%%%%%%%%%%%%%%%%%%%%%%%%%%%%%%%%%%%%%%%
\section{Dark energy parametrizations}
\label{sec:DE_param}
%%%%%%%%%%%%%%%%%%%%%%%%%%%%%%%%%%%%%%%%%%%%%%%%%%

\begin{figure}[htbp]
  \begin{center}
    \resizebox{50mm}{!}{
    \includegraphics[bb= 220 200 380 620,width=1\linewidth]{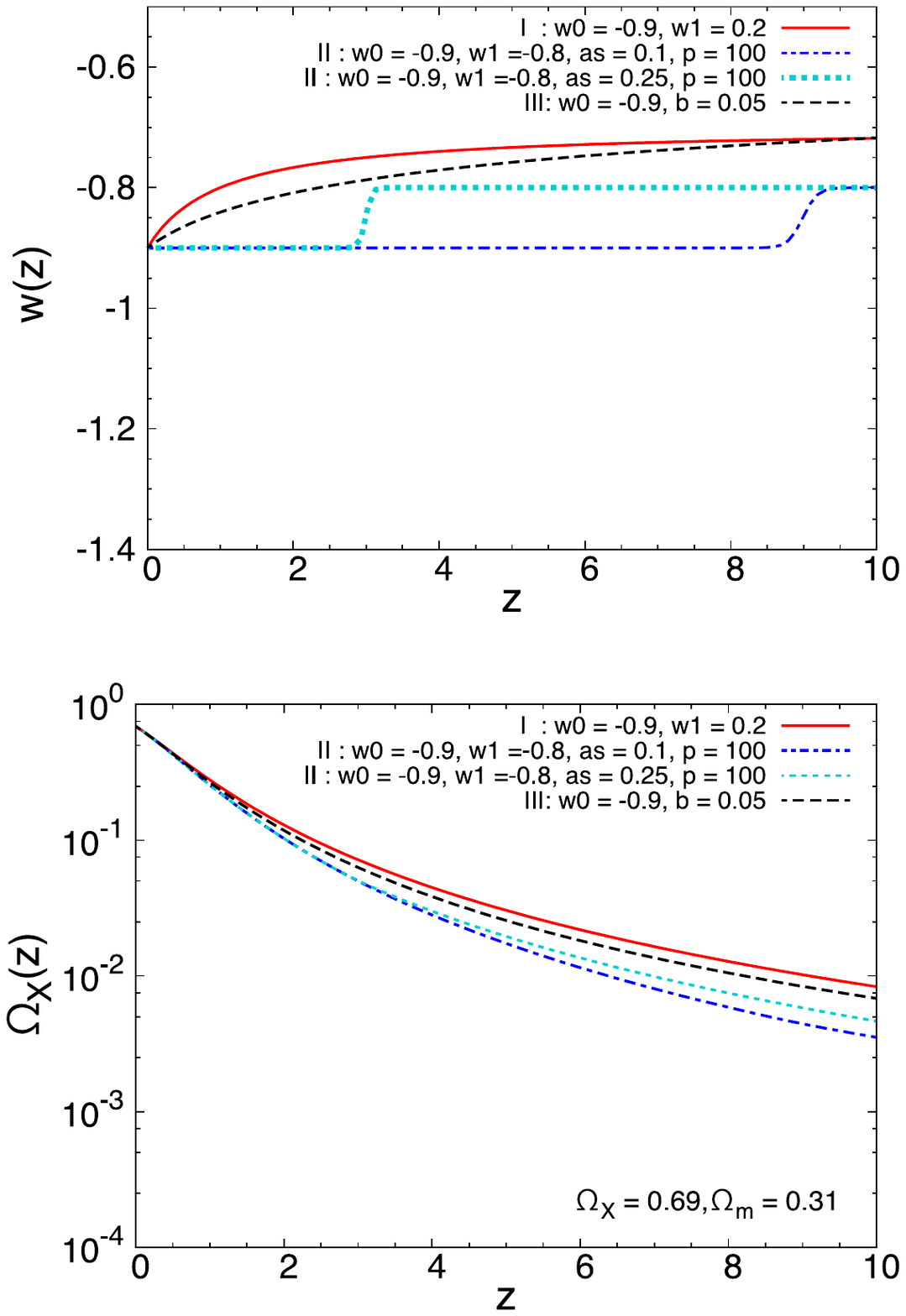}}
  \end{center}
%  \vspace{-40mm}
  \caption{
 Time evolutions of  
  EoS $w(z)$ (upper panel) and the density parameter $\Omega_X(z)$  (lower panel) 
  for the dark energy parametrizations considered in this paper.
  The labels I, II and III indicate the parameterization I, II and III, respectively.
  For the values of the parameters in each parametrization, we assumed the same ones as 
in Table. \ref{tab:summary} . 
  }
  \label{fig:EoS1}
\end{figure}

In this section, we summarize parametrizations for dark energy
equations of state $w_X$ which are used in our analysis. In
most models of dark energy, its EoS depends on time.
However, the time variation of $w_X$ is highly model-dependent, and
thus it is customary to assume some parametrization for $w_X$ to take
the time-dependence into account.  Although there have been proposed a
lot of possible parametrizations, we adopt some representative ones in
this paper.  For some other parametrizations, see, for example,
Ref.~\cite{Johri:2005qa}, in which various dark energy
parametrizations are discussed.
Behaviors of EoS $w_X(z)$ and the density parameter $\Omega_X(z)$
for the dark energy parametrizations adopted in this paper are shown in Fig.~\ref{fig:EoS1}.

\bigskip 
\noindent 
$\bullet$ {\bf Parametrization I}

One of simple ways to parametrize the time-varying $w_X$ is to assume
that $w_X$ varies linearly with the scale factor, which can be given
by the following form \cite{Chevallier:2000qy,Linder:2002et}:
\begin{equation}
\label{eq:EoS_1}
w_X (z) = w_0 + (1-a) w_1 = w_0 + \frac{z}{1+z} w_1,
\end{equation}
where $a$ and $z$ are respectively the scale factor and redshift\footnote{
Another parametrization in which $w_X$ changes linearly with the
redshift has also been adopted in some
literature~\cite{Huterer:2000mj,Weller:2001gf,Frampton:2002vv}.
However, here we do not consider such a parametrization.
}. 
$w_0$ represents the EoS at present while $w_0 + w_1$ is its value at far past.
With this parametrization, the energy density of dark energy can be given by
\begin{equation}
%\label{ }
\rho_X (z) = \rho_{X0} (1+z)^{3(1+w_0 + w_1)} \exp \left[ -\frac{3 w_1 z}{1+z} \right],
\end{equation}
with $\rho_{X0}$ being the dark energy density at present.

\bigskip 
\noindent 
$\bullet$ {\bf Parametrization II}

In some models of dark energy, its EoS suddenly changes
from some constant value to another.  This kind of changes cannot be
accommodated in the above parametrization.  The authors of
\cite{Hannestad:2004cb} have introduced a parametrization which can
represent this kind of sudden change of $w_X(z)$:
\begin{equation}
\label{eq:EoS_2}
w_X (z) 
= w_0 w_1 \frac{a^p + a_s^p}{w_1 a^p + w_0 a_s^p} 
= w_0 w_1 \frac{1 + \left( \frac{1+z}{1+z_s} \right)^p }{w_1 + w_0 \left( \frac{1+z}{1+z_s} \right)^p},
\end{equation}
where $a_s$ ($z_s$) corresponds to the scale factor (redshift) at which 
$w_X$ changes suddenly from $w_1$ to $w_0$. In the above expression, $p$
controls the width of the transition. 

The time evolution of the dark energy density in this parametrization is calculated as
\begin{equation}
%\label{ }
\rho_X (z) = \rho_{X0} (1+z)^{3(1+w_0)} \left( \frac{w_0 + w_1 (1+z_s)^p}{w_0 (1+z)^p + w_1 (1+z_s)^p} \right)^{3(w_0 - w_1)/p}.
\end{equation}

\bigskip 
\noindent 
$\bullet$ {\bf Parametrization III}

A commonly assumed dark energy evolves in such a way that its energy
density begins to dominate only at a late epoch of the Universe, and
it is subdominant at an earlier time. However, in some models such as
tracker quintessence model in which its EoS traces that of the
dominant component in the Universe (i.e., radiation or matter). Such a
model is sometimes called ``early dark energy'' since its energy
density contributes to the total one to some extent.  To represent
this kind of models, the following parametrization has been proposed
\cite{Wetterich:2004pv}:
\begin{equation}
\label{eq:EoS_3}
w_X (z) = \frac{w_0}{[1+b \log (1+z) ]^2},
\end{equation}
where $w_0$ corresponds to the present-day equation of state.  To
avoid too large value (or divergence) of $w_X$ at earlier times, $b$
should be positive.  Then, when $1+z$ is large, $w_X$ approaches to 0,
which is the same as that of matter.  Therefore, this kind of dark
energy can give some contribution at earlier times (in
matter-dominated epoch).  The energy density of dark energy of this
parametrization can be written as
\begin{equation}
%\label{ }
\rho_X (z) = \rho_{X0} (1+z)^{3(1+ \alpha_X(z))},
\end{equation}
where 
\begin{equation}
%\label{ }
\alpha_X (z) = \frac{w_0}{1 + b \log (1+z)}.
\end{equation}

%%%%%%%%%%%%%%%%%%%%%%%%%%%%%%%%%%%%%%%%%%%%%%%%%%
\section{Current constraints}
\label{sec:current_const}
%%%%%%%%%%%%%%%%%%%%%%%%%%%%%%%%%%%%%%%%%%%%%%%%%%

Before investigating the future expected constraints on dark energy
parameters, here we study constraints from current observations such
as Planck, BAO,  SNeIa, $H_0$ and weak lensing for each parametrization.  Although similar
analyses have been done in the literature for some parametrizations,
we update those constraints, and by doing those analyses, we can
choose fiducial values for the EoS parameters consistent with current
observational data in the next section.

We have performed Markov Chain Monte Carlo (MCMC) analyses by using a
modified version of CosmoMC \cite{Lewis:2002ah}, in which we have
accommodated the dark energy parametrizations introduced in
Section~\ref{sec:DE_param}.  For the sound speed $c_s^2$ which needs
to be specified when solving the perturbation equations, we fixed it
as $c_s^2=1$ in the analysis of this section. In addition to the
dark energy parameters, we have also varied the standard cosmological
parameters: baryon density $\Omega_bh^2$, dark matter density
$\Omega_{\rm DM}h^2$, the amplitude and the spectral index of the
scalar mode primordial power spectrum $A_s$ and $n_s$, the reionization optical depth $\tau$ and
the acoustic peak scale $\theta_s$.  Here $h$ is the normalized Hubble
parameter defined as $H_0 = 100h ~{\rm km/sec/Mpc}$.

In deriving constraints from current data, we adopt two different data
sets.  The first one is ``Planck+BAO+lensing'' which includes the
power spectra of the CMB temperature and polarization anisotropies
(TT+TE+EE at $\ell \ge 30$ and TT+TE+EE+BB at $\ell \le 29$) from
Planck~\cite{Aghanim:2015xee},  BAO scales in galaxy power
spectrum~\cite{Beutler:2011hx,Anderson:2013zyy,Ross:2014qpa} and the
Planck CMB lens power spectrum~\cite{Ade:2015zua}.  The other data set
we adopt in this analysis is ``Planck+BAO+lensing+$H_0$+SN+WL'' in which
data from  measurements of Hubble constant
$H_0=70.6 \pm 3.3$~km/s/Mpc~\cite{Efstathiou:2013via}, the JLA
compilation of type Ia supernovae~\cite{Betoule:2014frx} and the
CHFTLenS cosmic shear power spectrum~\cite{Heymans:2012gg} are added
to ``Planck+BAO+lensing.''

\begin{figure}[htbp]
  \begin{center}
    \resizebox{100mm}{!}{
    \includegraphics[bb= 228 316 384 475,width=1\linewidth]{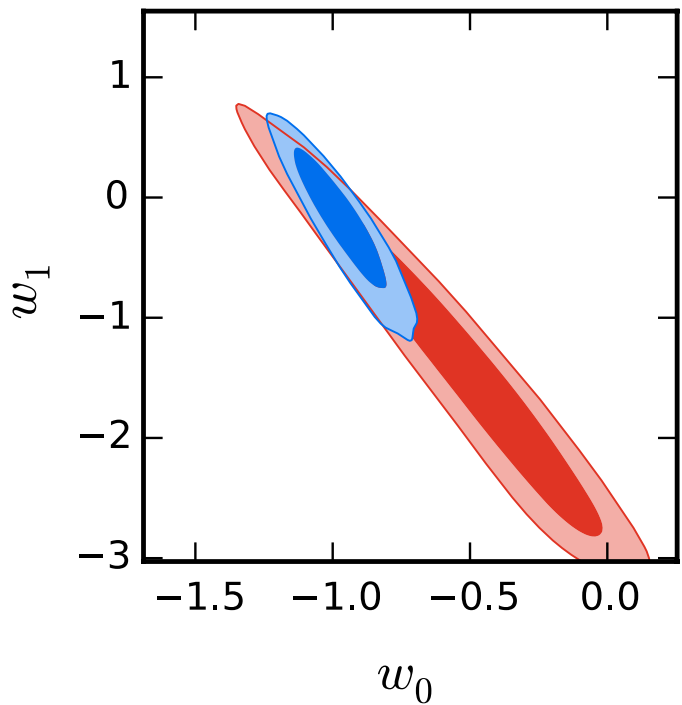}}
  \end{center}
%  \vspace{-40mm}
  \caption{Current constraint on the parametrization I in the
    $w_0$--$w_1$ plane.  We show the constraints from
    Planck+BAO+lensing (red region) and Planck+BAO+lensing+$H_0$+SNe+WL (blue
    region).  The dark and the light colors correspond to 
    1$\sigma$ and  2$\sigma$ allowed regions, respectively.  }
  \label{fig:CPL_current}
\end{figure}

\begin{figure}[htbp]
  \begin{center}
    \resizebox{170mm}{!}{
    \includegraphics[bb= 0 0 595 842,width=1\linewidth]{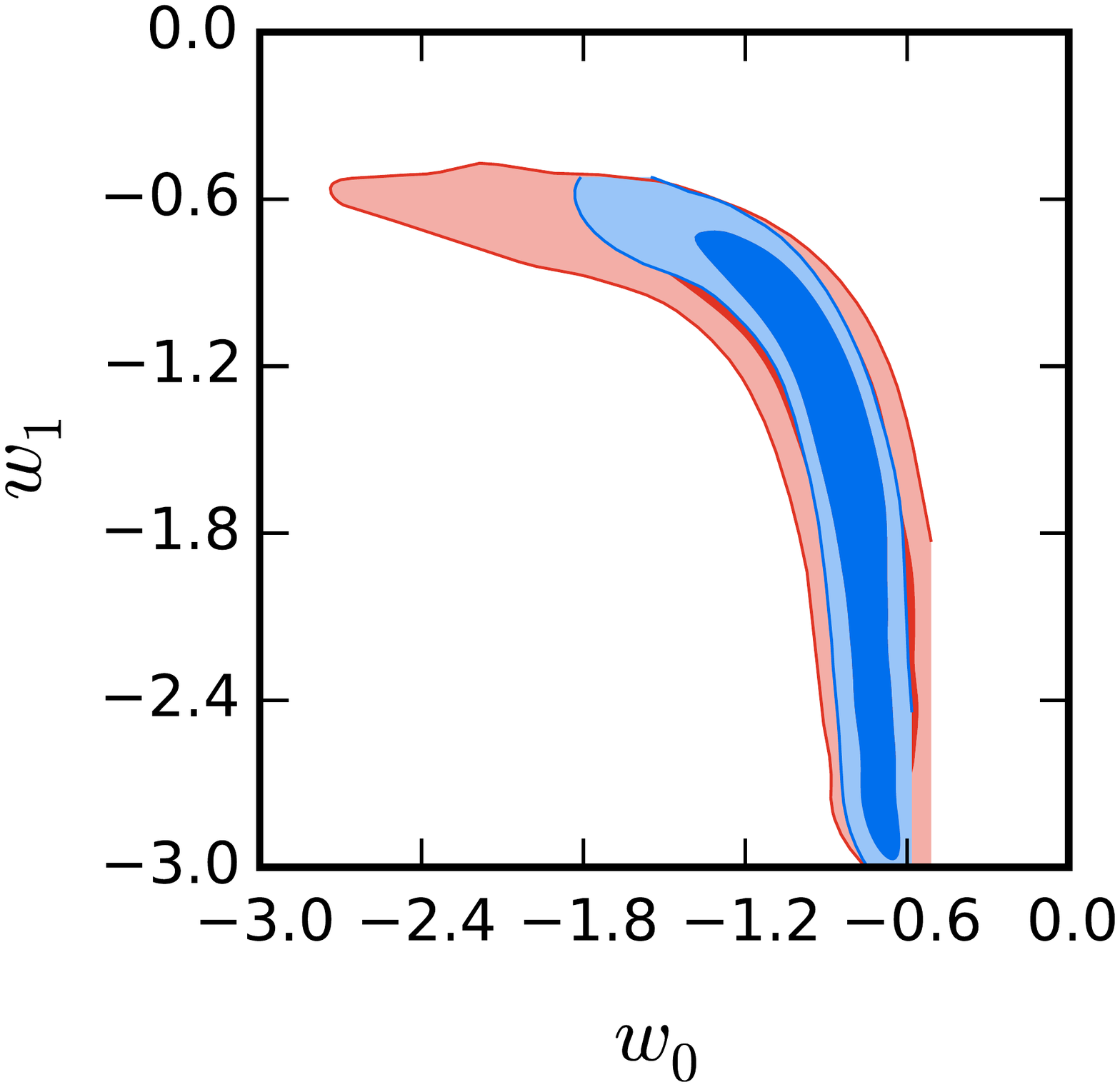}
    \includegraphics[bb= 0 0 595 842,width=1\linewidth]{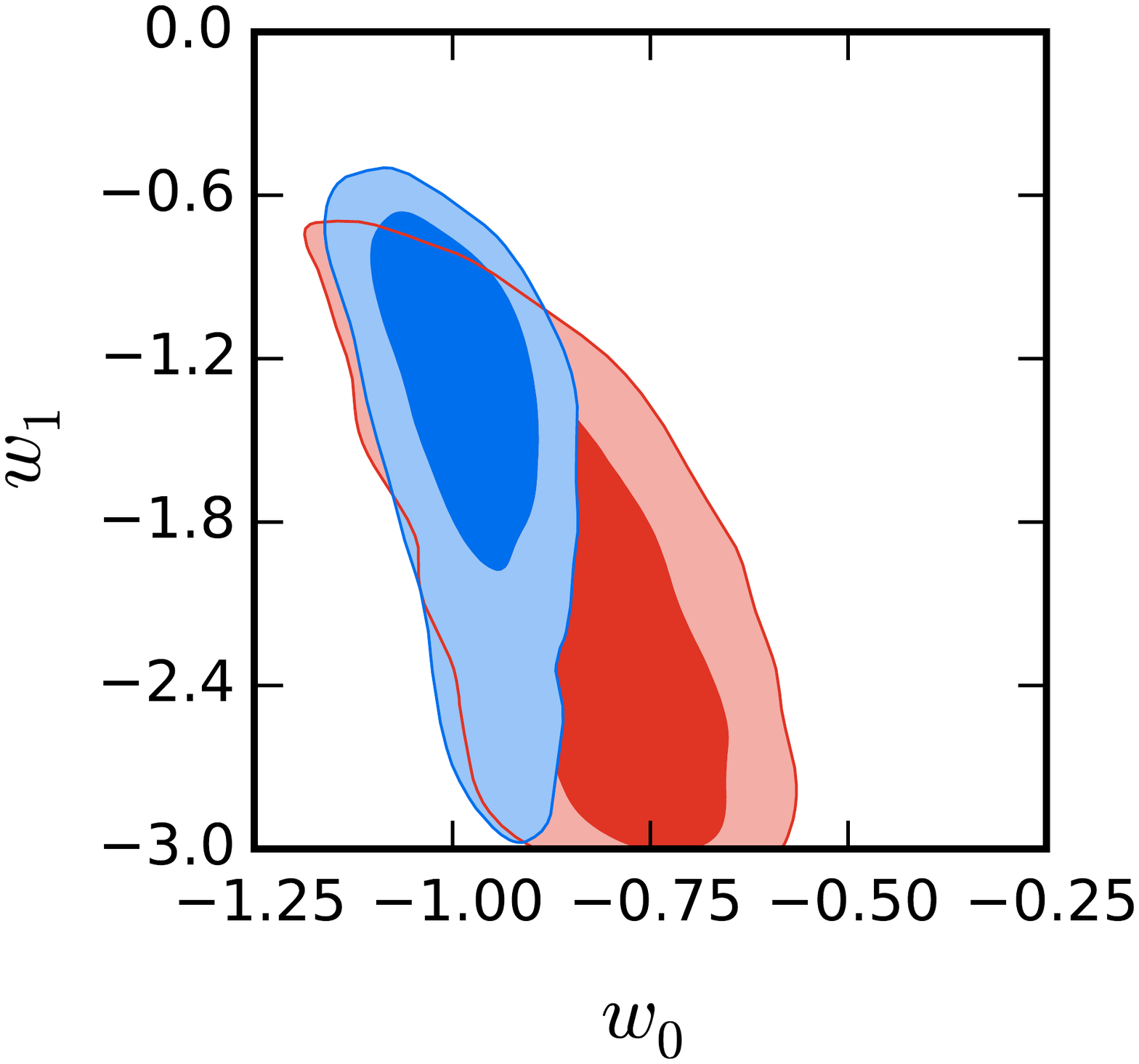}
}
  \end{center}
    \vspace{-20mm}
    \caption{The same as Fig.~\ref{fig:CPL_current}, but for the
      parametrization II in the $w_0$--$w_1$ plane.  We show the
      cases of $p=1$ at the left panel, and $p=100$ at the right
      panel, respectively. We chose $a_s=0.5$ commonly in the both
      panels.  }
  \label{fig:HM_current}
\end{figure}

\begin{figure}[htbp]
  \begin{center}
    \resizebox{100mm}{!}{
    \includegraphics[bb= 0 0 595 842,width=1\linewidth]{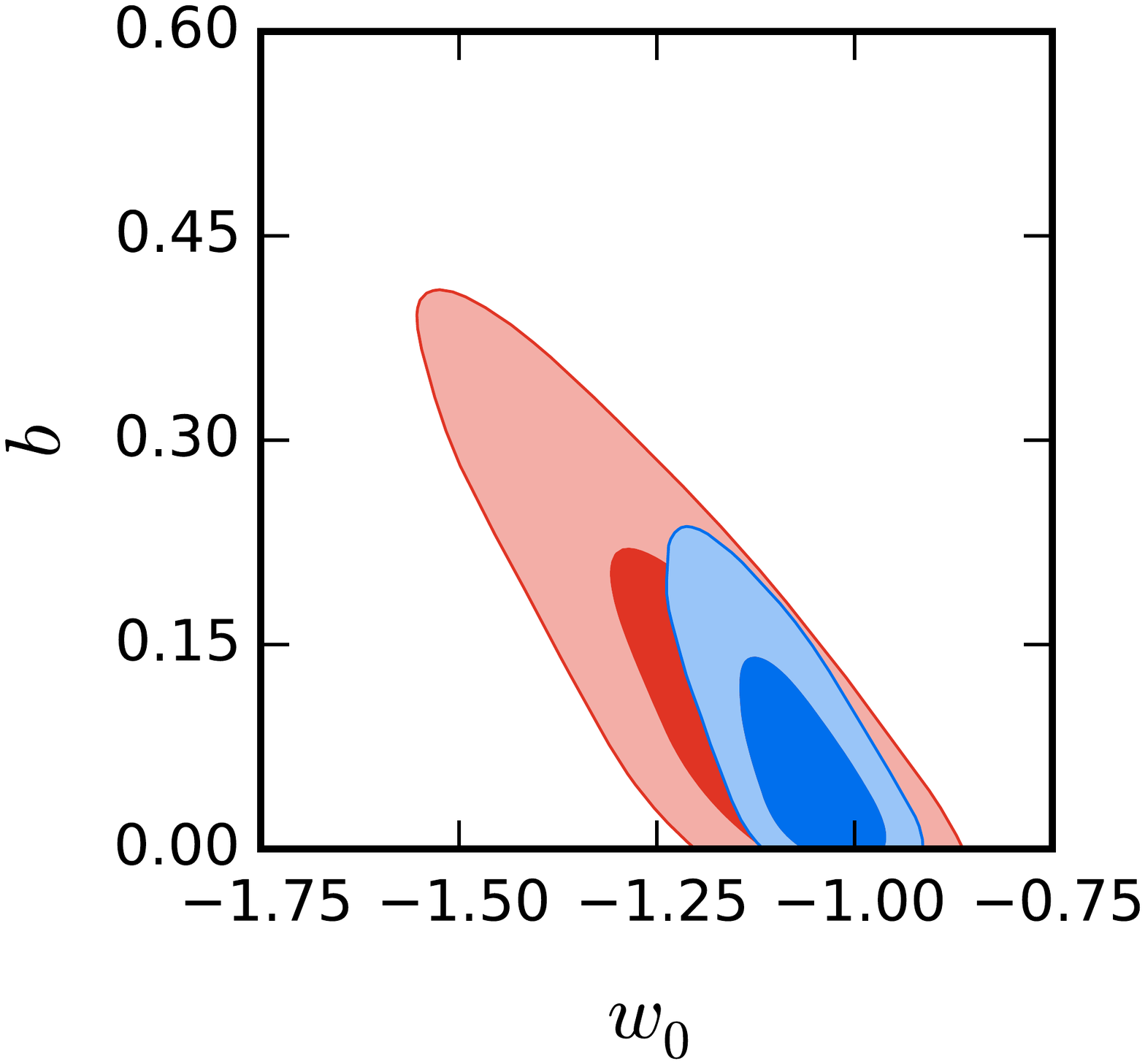}
}
  \vspace{-25mm}
  \end{center}
  \caption{The same as Fig.~\ref{fig:CPL_current}, but for the
    parametrization III in the $w_0$--$b$ plane.  }
  \label{fig:Wett_current}
\end{figure}

In Figs. \ref{fig:CPL_current}, \ref{fig:HM_current} and
\ref{fig:Wett_current}, we present the constraints for the
parametrizations I, II and III, respectively.  For the parametrization~I, 
we have assumed the prior for the EoS parameters $w_0$ and $w_1$ as
$ -3 \le w_0 \le 1$ and $ - 3 \le w_1 \le 3$, respectively.  The
constraint on this parametrization has been investigated in
\cite{Ade:2015rim}, using almost the same (but slightly different)
data sets.  Our result is consistent with the one obtained in
\cite{Ade:2015rim}.  For the parametrization~II, we have fixed $a_s$
and $p$ and show the constraints on the $w_0$--$w_1$ plane.  In
Fig.~\ref{fig:HM_current}, we present our results for the cases with
$(a_s, p)=(0.5,1)$ (left panel) and $(a_s, p)=(0.5, 100)$ (right
panel).  The prior for $w_0$ and $w_1$ are assumed as
$ -3 \le w_0 \le 1$ and $ - 3 \le w_1 \le 3$, as in the case for the
parametrization I.  The constraints on this parametrization has been
studied by using Planck and BAO in \cite{Jaber:2016ucq}.  For the
parametrization III, we have only two parameters for EoS, $w_0$ and
$b$. Therefore we show the constraint on the $w_0$--$b$ plane for this
parametrization. As stated in the previous section, this
parametrization captures the property of the so-called early dark
energy models.  In \cite{Ade:2015rim}, the early dark energy model
been studied, but with a different parametrization from the one adopted
here.  Yet another parametrization of early dark energy has also been
discussed in \cite{Pu:2014goa} by using Planck 2013 data.

Having obtained the current constraints on the parameters for EoS, we choose
fiducial values of the EoS parameters, which are allowed within at
2$\sigma$ level, in order to perform the Fisher matrix analysis in the
next section.

%%%%%%%%%%%%%%%%%%%%%%%%%%%%%%%%%%%%%%%%%%%%%%%%%%
\section{Future expected constraints}
\label{sec:future_const}
%%%%%%%%%%%%%%%%%%%%%%%%%%%%%%%%%%%%%%%%%%%%%%%%%%

Now in this section, we present the expected constraints from future
21 cm experiments such as SKA and Omniscope in combination with future
CMB observations such as COrE+.  For comparison, we also show the results with the specification of Planck. 
In addition, we combine future observations of SNe, BAO and direct Hubble constant measurement to
obtain the expected constraints.  For this purpose, we adopt the Fisher matrix analysis. 
The specifications of observations assumed  in this paper are summarized in Appendix~\ref{sec:spec}.
Since the method of the Fisher analysis for
the 21cm fluctuations and CMB adopted in this paper is the same as the
one in our previous paper \cite{Kohri:2013mxa}, 
here we just briefly summarize them. 
For the details, we refer the readers to \cite{Kohri:2013mxa,Mao:2008ug}. 
We also briefly describe our methods for the analysis of future observations of SNe, BAO and the direct Hubble constant measurement.
After summarizing our methods for the Fisher analysis, we present our results on the expected constraints on dark energy parameters for
each parametrization in order.

In our Fisher analysis, the fiducial values of the cosmological parameters are set to the
mean values from the analysis of Planck
TT,TE,EE+lowP+lensing+$H_0$+SNe+BAO \cite{Ade:2015xua}:
$\Omega_bh^2 = 0.0223, \Omega_mh^2 =0.1417, \Omega_{X}=0.6911,
\tau = 0.066, A_s 10^{10} = 21.42$ and $n_s = 0.9667$, and $Y_p = 0.25$,
where $Y_p$ is the primordial $^4\textrm{He}$ mass fraction.
In our analysis, the total mass of neutrinos is fixed as $\Sigma m_{\nu}=0.06$~eV, and we assume that the hierarchy of the masses is normal ordering.  
For the EoS parameters, we choose multiple sets for each parametrization.

%%%%%%%%%%%%%%%%%%%%%%%%%%%%%%%%%%%%%%%%%%%%%%%%%%%
\subsection{21cm}
\label{subsec:21cm_Fisher}
%%%%%%%%%%%%%%%%%%%%%%%%%%%%%%%%%%%%%%%%%%%%%%%%%%%%%%%

The Fisher matrix for 21 cm fluctuations is given by 
\begin{equation}
%\label{ }
F_{ij}^{\rm (21cm)} 
= 
\sum_{\rm pixels} \frac{1}{\left( \delta P_{21} ({\bm u}) \right)^2} 
\frac{\partial P_{21} ({\bm u})}{\partial p_i} \frac{\partial P_{21} ({\bm u})}{\partial p_j},
\end{equation}
where $P_{21} ({\bm u})$ and $\delta P_{21} ({\bm u})$ are the 21 cm power spectrum  and its error in ${\bm u}$ space, respectively, 
and $p_i (p_j)$ represents cosmological parameters.
Here ${\bm u}$ is the Fourier dual of 
\begin{equation}
%\label{ }
{\bm \Theta} = \theta_x \hat{e}_x + \theta_y \hat{e}_y + \Delta f \hat{e}_z = {\bm \Theta}_\perp + \Delta f \hat{e}_z,
\end{equation}
where ${\bm \Theta}_\perp $ is the measured angular position on the sky and $\Delta f$ represents the frequency
which is given by the difference from the central redshift of a given bin. 
By adopting the flat-sky approximation, ${\bm u}$ can be related to ${\bm k}$ which is the Fourier dual to the position vector ${\bm r}$ as
\cite{Kohri:2013mxa,Mao:2008ug}
\begin{equation}
%\label{ }
{\bm u}_\perp = d_A (z) {\bm k}_\perp, 
\qquad
u_\| = y (z) k_\|, 
\end{equation}
where $d_A(z)$ is the angular diameter distance to the redshift $z$ and $y(z) = \lambda_{21} (1+z)^2 /H(z)$ with $\lambda_{21}$ being the wavelength of 
21 cm line. A vector with the subscript $\perp$ ($\|$)  denotes   the component  perpendicular (parallel)  to the line of sight. From the above formulas, 
the power spectra in ${\bm u}$-space and ${\bm k}$-space are related as 
\begin{equation}
%\label{ }
P_{21} ({\bm u}) = \frac{1}{d_A (z)^2 y (z)} P_{21} ({\bm k}).
\end{equation}

The power spectrum $P_{21} ({\bm k})$ is defined by 
\begin{equation}
\label{eq:power_def}
\left\langle \delta T_b ({\bm k}) \delta T_b^{\ast} ({\bm k'}) \right\rangle = (2\pi)^3 \delta^{(3)} ({\bm k} - {\bm k'}) P_{21} ({\bm k}),
\end{equation}
where $\delta T_b ({\bm k})$ is fluctuations of the differential brightness temperature of the 21 cm line ($T_b$) relative to that of CMB ($T_{\rm cmb}$).
$T_b$ in the position space is given by 
\begin{equation}
%\label{ }
T_b ({\bm x}) = \bar{T}_b \left( 1 - \bar{x}_i ( 1 + \delta_x ({\bm x}) ) \right) (1 + \delta ({\bm x}) )\left( 1 - \frac{1}{aH} \frac{d v_r}{dr} \right),
\end{equation}
where  barred quantities represent their averaged ones, and here we have assumed $T_s \gg T_{\rm cmb}$ with $T_s$ the spin temperature
since we consider the reionization era in this paper. $\bar{T}_b$ is  the average brightness temperature, which can be written as
\begin{equation}
\bar{T}_b  \simeq 27 \bar{x}_{H} 
\left( \frac{1-Y_p}{1-0.25} \right)
\left( \frac{\Omega_bh^2}{0.022} \right) \left( \frac{0.14}{\Omega_m h^2} \frac{1+z}{10} \right)^{1/2} ~{\rm mK},
\end{equation}
where $x_H$ is the neutral hydrogen fraction. 
$\delta_x$ and $\delta$ are  fluctuations of the ionization fraction 
$x_i= 1 - x_H$
and the hydrogen number density $n_H$, respectively.
$(1/aH) (d v_r/dr)$ represents the peculiar velocity and can be treated as a perturbation, which we define to be $\delta_v ({\bm x}) \equiv (1/aH) (d v_r/dr)$.
The Fourier transform of $\delta_v ({\bm x}) $ is related to $\delta ({\bm k})$ as  $\delta_v ({\bm k}) = -\mu^2 \delta ({\bm k})$ with 
$\mu = \hat{\bm k}\cdot \hat{\bm n}$ being the cosine of the angle of ${\bm k}$ relative to the line of sight. 
Since the fluctuations of $T_b$ is given by 
$\delta T_b ({\bm x}) = T_b ({\bm x}) - \bar{T}_b $, 
the power spectrum in the ${\bm k}$-space can be written as 
\begin{equation}
%\label{ }
P_{21} ({\bm k}) 
= 
\left[  {\cal P}_{\delta\delta} (k) - 2 {\cal P}_{x\delta} (k) + {\cal P}_{xx}(k) \right]  
+ \mu^2 \left[ 2 {\cal P}_{\delta\delta} (k)- 2 {\cal P}_{x\delta} (k)\right]
+\mu^4  {\cal P}_{\delta\delta} (k),
\end{equation}
where $k = |{\bm k}|$, 
${\cal P}_{\delta\delta} = \bar{T}_b^2 P_{\delta\delta}, 
{\cal P}_{x\delta} = \bar{T}_b^2 ( \bar{x}_i /\bar{x}_H )P_{x\delta}
$ and 
$ {\cal P}_{xx} = \bar{T}_b^2 (\bar{x}_i^2 /\bar{x}_H^2 ) P_{xx}$ 
with $P_{\delta\delta}, P_{x\delta}$ and $P_{xx}$ being the power
spectra for $\delta$ and $\delta_x$ defined as the same as that for
$P_{21}$ given in Eq.~\eqref{eq:power_def}.  Here $P_{\delta\delta}$
represents the matter fluctuations which carries the information of
cosmology, and hence can probe the nature of dark energy.
$P_{x\delta}$ and $P_{xx}$ are the power spectra involving
$\delta_x$. After the reionization has started, these power spectra
can significantly affect the total 21 cm power spectrum, and they are
determined by the physics of reionization.  For these power spectra,
here we assume the following forms which can be well fitted by a
result from radiative transfer simulations
of~\cite{McQuinn:2006et,McQuinn:2007dy}:
\begin{eqnarray}
%\label{ }
{\cal P}_{xx} (k) &=& b_{xx}^2 \left[ 1 + \alpha_{xx} (kR_{xx}) + \left( k R_{xx} \right)^2 \right]^{-\gamma_{xx} /2} {\cal P}_{\delta\delta} (k), \\
{\cal P}_{x\delta} (k) &=& b_{x\delta}^2  e^{ - \alpha_{x\delta} (kR_{x\delta}) - \left( k R_{x\delta} \right)^2 } {\cal P}_{\delta\delta} (k). 
\end{eqnarray}
The amplitudes and the shapes of the spectra are described  by the parameters $b_{xx}, b_{x\delta}, \alpha_{xx}, \gamma_{xx}$ and $\alpha_{x\delta}$,
and  $R_{xx}$,  $R_{x\delta}$ are the parameters characterizing the effective size of the  ionized bubbles.
In our Fisher analysis, we also vary these ionization parameters, but marginalize over them to obtain expected constraints on 
dark energy parameters.  For the fiducial values,  we assume the same ones which are given 
in Table III of \cite{Mao:2008ug} (or Table 1 of \cite{Kohri:2013mxa}).
In our analysis, we divide the redshift bins into four: $z= 6.75 - 7.25, 7.25 - 7.75, 7.75 - 8.25$ and $8.25 - 10.05$.

The error power spectrum $\delta P_{21} ({\bm u})$ is given by 
\begin{equation}
%\label{ }
\delta P_{21} ({\bm u}) = \frac{P_{21} ({\bm u}) + P_N (u_\perp) }{N_c^{1/2}},
\label{eq:variance_P21}
\end{equation}
where $P_N (u_\perp)$ is the noise power spectrum, and the first term
in the RHS represents the sample variance.  The quantities to describe
$P_N$ are summarized in Table~\ref{tab:noise_power}.  In
order to avoid the foreground contamination, we do not use the
wavelength less than $k_{{\rm min} \|} = 2\pi / (yB)$\footnote{
Although we simply remove the modes with this criterion, the foreground could leak to the so-called ``foreground wedge," 
which can be a serious problem in deriving the constraints \cite{Datta:2010pk,Vedantham:2011mh,Morales:2012kf,Parsons:2012qh,Trott:2012md,Thyagarajan:2013eka,Hazelton:2013xu,Liu:2014bba,Liu:2014yxa,Thyagarajan:2015ewa}.
To estimate the impact of the foreground wedge on our results, we have also made the analysis by removing the modes 
with $ \mu < \mu_{\rm min} =  k_\| / \sqrt{k_\perp^2 + k_\|^2}$ \cite{Seo:2015aza}, where $\mu_{\rm min}$ can be $\sim 0.95$ at $z \sim 8$.
For such a value of $\mu_{\rm min}$, we found that the error can become larger by a bit less than a factor of 2  in some particular cases. However, 
we note that even in such a case,
the constraints can be improved by adding the information of 21cm.
},
where $B$ is the band width.
We also cut the wavelength larger than $k_{\rm max} = 2~{\rm Mpc}^{-1}$ not to be
affected by non-linear effect.

\bigskip
\begin{table}[h]
\begin{tabular}{rl}
\hline \hline 
 Noise power spectrum: & $P_N (u_\perp) = \left( \displaystyle\frac{\lambda^2(z) T_{\rm sys} (z) }{A_e (z)} \right)^2 \displaystyle\frac{1}{t_0 n (u_\perp)} $  \\    
 System temperature: & $T_{\rm sys} = T_{\rm sky} + T_{\rm rcvr}$ \\
 Sky temperature: & $T_{\rm sky} = 60 \left( \lambda / [m] \right)^{2.55}$ \\
 Receiver noise: & $ T_{\rm rcvr} = 0.1 T_{\rm sky} + 40 [{\rm K}]$ \\
 Number of independent cells: & $N_c = 2\pi k_\perp \Delta k_\perp \Delta k_\| V(z) / (2\pi)^3$    \\
 Survey volume:  &$ V(z) = d_A(z)^2 y(z) B \times {\rm FoV}$  \\
 Effective collecting area: & $A_e$ \\
 Observation time: & $t_0$ \\
 Number density of the baseline: & $n (u_\perp)$ \\
 Field of view: & FoV \\
\hline \hline
\end{tabular}
  \centering 
  \caption{Quantities to describe the error power spectrum $P_N$. }
  \label{tab:noise_power}
\end{table}

The methodology described above is basically the same as the one
adopted in \cite{Kohri:2013mxa}. Since the publication of
\cite{Kohri:2013mxa} however, the specification of SKA has been
changed. Hence we adopt the new specification and summarize it in
Table~\ref{tab:SKA_spec} in Appendix~\ref{sec:spec}.

We also note here that, in the following analysis, we focus on the 21 cm signals of  redshift ranges 
$6.8 < z < 10$, which corresponds to the era of reionization,  and
investigate how such high redshift information can be useful to probe the nature of dark energy.
Although 21 cm signals at low redshifts after reionization would also be very helpful \cite{Loeb:2008hg,Wyithe:2008mv,Visbal:2008rg}, in this paper
we focus on the 21 cm signals only at high redshifts.

%%%%%%%%%%%%%%%%%%
\paragraph{Number density of the baseline:}
%%%%%%%%%%%%%%%%%%
%
To set the number density of the baseline  $n(u_\perp)$, 
we assume an azimuthally symmetric distribution for the antenna stations and 
adopt the following density profile which is consistent with the specification of the originally planned SKA1 \cite{Kohri:2013mxa}:
\begin{align}
\rho(r) = 
\left\{
\begin{array}{lll}
 \rho_{0}r^{-1},   &\rho_{0} \equiv \frac{13}{16\pi\left(\sqrt{10}-1\right) }  \ {\rm m}^{-2}
& \hspace{60pt} r \leq 400 \ {\rm m},\\
 \rho_{1}r^{-3/2}, &\rho_{1} \equiv \rho_{0} \times 400^{1/2}, & \ \ \ 400 \ {\rm m} < r \leq 1000 \ {\rm m}, \\
 \rho_{2}r^{-7/2}, &\rho_{2} \equiv \rho_{1} \times 1000^{2}, & \ \ 1000 \ {\rm m} < r \leq 1500 \ {\rm m}, \\
 \rho_{3}r^{-9/2}, &\rho_{3} \equiv \rho_{2} \times 1500 ,          & \ \ 1500 \ {\rm m} < r \leq 2000 \ {\rm m}, \\
 \rho_{4}r^{-17/2},&\rho_{4} \equiv \rho_{3} \times 2000^{4}, & \ \ 2000  \ {\rm m} < r \leq 3000 \ {\rm m}, \\
\end{array}
\right.
\label{eq:density_profile}
\end{align}
where $r$ is a radius from the center of the array. 
95\% of the stations are assumed to be in the region with $r < 3000~{\rm m}$.

Although the original plan of SKA1 has assumed 911 antenna stations, 
in our analysis, we take the number of stations consistent with the re-baseline design of SKA 
in which $N_{\rm ant} = 911 /2$ for SKA1 and $N{\rm ant} = 911\times  4$ for SKA2 
as shown in Table~\ref{tab:SKA_spec}. Therefore the number density of the baseline  $n(u_\perp)$
can be evaluated as 
\begin{align}
n_{{\rm SKA1}}(u_\perp) 
&= n_{{\rm origSKA1}}(u_\perp) \times \left( \frac{1}{2} \right)^{2}, \\
n_{{\rm SKA2}}(u_\perp)
&= n_{{\rm origSKA1}}(u_\perp) \times 4^{2},
\end{align}
where $n_{{\rm origSKA1}}(u_\perp)$ is the number density of the baseline for the original design of SKA1 which is 
evaluated by its density profile of Eq.~\eqref{eq:density_profile}. 
$n_{{\rm SKA1}}(u_\perp)$ and $n_{{\rm SKA2}}(u_\perp)$
are the ones for  SKA1 and SKA2, respectively.
For Omniscope, we assume the same one as Ref. \cite{Mao:2008ug}.

%%%%%%%%%%%%%%%%
%%%%%%%%%%%%%%%% 

%%%%%%%%%%%%%%%%%%%%%%%%%%%%%%%%%%%%%%%%%%%%%%%%%%%
\subsection{CMB}
\label{subsec:CMB_Fisher}
%%%%%%%%%%%%%%%%%%%%%%%%%%%%%%%%%%%%%%%%%%%%%%%%%%%%%%%

The Fisher matrix of CMB is written as
\begin{equation}
%\label{ }
F_{ij}^{\rm (CMB)}  = \sum_l \left( \frac{2l+1}{2} \right) f_{\rm sky} {\rm Tr}  \left[ {\bf C}_l^{-1}  \frac{\partial {\bf C}_l}{\partial p_i}  {\bf C}_l^{-1}  \frac{\partial {\bf C}_l}{\partial p_j} \right],
\end{equation}
where $p_i (p_j)$ represents cosmological parameters, and ${\bf C}_l $ is the covariance matrix of CMB which is given by 
\begin{equation}
%\label{ }
 {\bf C}_l =
\begin{pmatrix}
C_l^{TT} + N_l^T  &  C_l^{TE}  & C_l^{Td}   \\
C_l^{TE}  &   C_l^{EE} + N_l^P  & 0 \\
C_l^{Td}  & 0   & C_l^{dd} + N_l^d 
\end{pmatrix},
\end{equation}
where $C_l^X$ is the angular power spectrum for unlensed CMB  ($X = TT, TE, EE$),   weak lensing deflection angle field ($X=dd$)
and   cross correlation between $TT$ and $d$  ($X=Td$).
$N_l^Y$ is the noise power spectrum which is given by
\begin{equation}
%\label{ }
N_l^Y  (\nu) = \Delta_Y^2 \exp \left[ l(l+1) \sigma_b^2 (\nu) \right],
\end{equation}
where $\Delta_Y$ is the experimental noise, and $\sigma_b = \theta_{\rm FWHM}/\sqrt{8 \ln 2}$ represents 
the beam width.  When the multiple frequency channels are used, the total noise power spectrum can be provided by
\begin{equation}
%\label{ }
\left( N_l^Y \right)^{-1}  = \sum_{\nu_i} \frac{1}{N_l^Y (\nu_i)}.
\end{equation}
For the noise power spectrum for weak lensing deflection angle $N_l^d$, we use FUTURCMB code \cite{futurcmb} which adopts the quadratic estimator for lensing reconstruction \cite{Okamoto:2003zw}.
$f_{\rm sky}$ is the fraction of the sky measured and we assume $f_{\rm sky}=0.65$.

For the specifications of CMB, we assume COrE+ and Planck. 
Although the specification of COrE+ can be found  in \cite{COrE+-proceeding}, 
it is considered to be the one at the planning stage.
Therefore, we use the values from Ref.~\cite{core+}, which is the original specification of
COrE.  Although the specification adopted in this work is a 
little bit different from the one assumed in \cite{Kohri:2013mxa}, it
does not change our results at all. We tabulate its specification in Table~\ref{tab:core_spec}.  
For Planck, we assume the same specification as that in \cite{Kohri:2013mxa}.

%%%%%%%%%%%%%%%%%%%%%%%%%%%%%%%%%%%%%%%%%%%%%%%%%%%
\subsection{BAO, SNe and Hubble constant}
\label{subsec:BAO_SNFisher}
%%%%%%%%%%%%%%%%%%%%%%%%%%%%%%%%%%%%%%%%%%%%%%%%%%%%%%%

Here we describe our method of the analysis to obtain joint constraints from CMB,
21cm fluctuations, BAO, SNe and direct measurements of the Hubble
constant $H_0$.  As mentioned above, the methods to analyze CMB and 21
cm fluctuations are the same as the one adopted in
\cite{Kohri:2013mxa}. However, future observations of BAO, SNe and
$H_0$ were not included in \cite{Kohri:2013mxa}, and therefore, before
showing our results on future constraints, we summarize formalisms of
our analysis for BAO, SNe and $H_0$ (for reference, see also
Refs.\cite{Oyama:2015gma,Oyama:2015vva,Oyama:2016lor}).

%%%%%%%%%%%%%%%%%%%%%%%%%%%%%%
\subsubsection[BAO]{BAO}
\label{subsubsec:BAOFisher}
%%%%%%%%%%%%%%%%%%%%%%%%%%%%%%

Observations of BAO can probe the comoving angular diameter distance $d_{A}(z)$ and the Hubble parameter $H(z)$,
which depend on underlying cosmological models. Therefore they are also affected by the EoS of dark energy. 
For our analysis,  $\ln(d_{A}(z))$ and $\ln(H(z))$ are treated as observables and the Fisher matrix for BAO 
is given by 
\begin{eqnarray}
F^{({\rm BAO})}_{\alpha, \beta} 
&=& 
F^{({\rm BAO}), d}_{\alpha, \beta} + F^{({\rm BAO}), H}_{\alpha, \beta} \notag \\
&=&
\sum_{i} \frac{1}{\sigma_{d}^2(z_{i})} 
\frac{\partial \ln (d_A (z_i)) }{\partial \theta_{\alpha}}
\frac{\partial \ln (d_A (z_i)) }{\partial \theta_{\beta}}
+ 
\sum_{i} \frac{1}{\sigma_{H}^2(z_{i})} 
\frac{\partial \ln (H (z_i)) }{\partial \theta_{\alpha}}
\frac{\partial \ln (H(z_i)) }{\partial \theta_{\beta}},
\label{eq:FBAO}
\end{eqnarray}
where the sum should be performed for the redshift bins whose width, and its central value for the $i$-th bin are  denoted as $\Delta z_i$ and $z_i$, respectively\footnote{
In general, measurements of $d_A$ and $H$ from BAO are correlated, 
and it should give rise to off-diagonal components in the covariant matrix that 
are omitted in Eq.~\eqref{eq:FBAO}.
According to an analytical estimation~\cite{Seo:2007ns}, the correlation coefficients can be 
as large as 0.4. However, we find that 
the impact of the off-diagonal components at this level 
on the constraints of cosmological parameters are fairly small (only about 4-5\%) in our results.
}.
$\sigma_{d}(z_{i})$ and $\sigma_{H}(z_{i})$ are errors of $\ln(d_{A}(z_{i}))$ and $\ln(H(z_{i}))$ at each redshift bin.
$\theta_\alpha$  (and $\theta_\beta$) represents the cosmological parameters which are relevant to observations of BAO:
$\Omega_{m}h^2$,  $\Omega_{X}$\footnote{
We can also take $h$ and $\Omega_{X}$ as the primary parameters for the analysis when we assume a flat Universe. 
However, in our analysis, we treat $\Omega_{m}h^2$ and $\Omega_{X}$ as the primary parameters.
} and the dark energy EoS parameters.
The component $F_{\alpha,\beta}$ which are irrelevant to the BAO observables are set to be zero.

In the following analysis, for a future BAO observation, we adopt the specification of Dark Energy
Spectroscopic Instrument (DESI)~\cite{DESI:web,Font-Ribera:2013rwa}. 
We assume the redshift range, bins and expected errors for  $\ln(d_{A}(z_{i}))$ and $\ln(H(z_{i}))$
as the same as those given in Table 5 of \cite{Font-Ribera:2013rwa}.

%%%%%%%%%%%%%%%%%%%%%%%%%%%%%%
\subsubsection[H0]{Direct measurement of $H_0$ }
\label{subsubsec:H0Fisher}
%%%%%%%%%%%%%%%%%%%%%%%%%%%%%%

In future, direct measurement of the Hubble constant $H_0$ will be more precise
to reach the level of $1~\%$
\cite{Macri:2006wm,Argon:2007ry,Greenhill:2009yi}.  Here we assume
that we can determine the Hubble constant at a $1~\%$ accuracy in our
analysis, which is expressed in the Fisher matrix to
be~\cite{Wu:2014hta}
\begin{eqnarray}
\label{eq:H_fisher}
F^{(H_0)}_{\alpha, \beta} 
= \frac{1}{ (0.01 \times h_{\rm fid})^2}  \delta_{\alpha, h} \delta_{\beta, h}.
\end{eqnarray}
The Kronecker delta symbols indicate that the Fisher matrix component of $(h, h)$ is the only nonzero entity in the Fisher matrix.

However, in fact, in our analysis, we do not adopt $h$ as a primary parameter, but it is derived from other cosmological parameters. 
Hence we need to convert the Fisher matrix Eq.~\eqref{eq:H_fisher} to that for our primary cosmological parameters. 
Among the primary parameters adopted in our analysis, $\Omega_mh^2$ and $\Omega_X$ are relevant to $h$, 
and therefore, we transform the Fisher matrix with $(\alpha, \beta) = (h, \Omega_X)$ to the one with  $(\alpha, \beta) = (\Omega_mh^2, \Omega_X)$ 
by using the following transformation \cite{Albrecht:2006um}:
\begin{eqnarray}
\tilde{F}_{l,m} = 
\frac{\partial \theta_{j}}{\partial \tilde{\theta}_{l}}
\frac{\partial \theta_{k}}{\partial \tilde{\theta}_{m}}
F_{jk}, 
\end{eqnarray}
from which we obtain 
\begin{align}
\tilde{F}^{(H_0)}_{\alpha, \beta} = 
\left(
\begin{array}{cc}
\tilde{F}_{\Omega_{m}h^2, \Omega_{m}h^2} &
\tilde{F}_{\Omega_{m}h^2, \Omega_{X}} \\
\tilde{F}_{\Omega_{m}h^2, \Omega_{X}} & 
\tilde{F}_{\Omega_{X}, \Omega_{X}}
\end{array}
\right)
=
\frac{1}{(0.01 \times h_{\rm fid})^2}
\left(\frac{1}{2\Omega_{m}h^2}\right)^2
\left(
\begin{array}{cc}
h^2 & h^4 \\
h^4 & h^6
\end{array}
\right).
\end{align}
We use this Fisher matrix to incorporate a future observation of direct measurement of the Hubble constant 
in the following analysis.

%%%%%%%%%%%Super Nova%%%%%%%%%%%%%%%%%%%%%%%%%%%%%%%%%%
\subsubsection{Supernovae}
\label{subsubsec:SNFisher}
%%%%%%%%%%%%%%%%%%%%%%%%%%%%%%%%%%%%%%%%%%%%%%%%%%%%%%%

Here we describe our method of the Fisher matrix analysis for SNe. 
For details, we refer the readers to \cite{Albrecht:2006um,Frieman:2008sn}.
Observations of SNe probe the apparent magnitude $m$ for each SNe which is related to the absolute magnitude $M$ by 
the following relation:
\begin{equation}
m = M + \mu(z),
\end{equation}
where $\mu(z)$ is the distance modulus and is related to the luminosity distance $d_{L}(z)$ as follows:
\begin{equation}
\mu(z) = 5\log_{10}\left(\frac{d_{L}(z)}{{\rm Mpc}}\right) + 25.
\end{equation}
In a flat Universe, the luminosity distance is given by
\begin{equation}
d_{L}(z) = (1+z) \int^{z}_{0} \frac{d\bar{z}}{H(\bar{z})}.
\end{equation}
Since the mean apparent magnitudes $m$ of each bin are the observable, the Fisher matrix for SNe  is written as 
\begin{align}
F^{({\rm SN})}_{\alpha, \beta} &=
\sum_{i} \frac{1}{\sigma_{\textrm{tot},i}^2} 
\frac{\partial m(\bar {z}_{i})}{\partial \theta_{\alpha}}
\frac{\partial m(\bar {z}_{i})}{\partial \theta_{\beta}}, \\
\sigma_{\textrm{tot},i}&=\sqrt{\sigma_{\textrm{stat},i}^2+\sigma_{\textrm{sys},i}^2},
\end{align}
where $\bar{z}_i$ is the mean redshift for the $i$-th bin and
$\sigma_{\textrm{tot},i}$, $\sigma_{\textrm{stat},i}$ and $\sigma_{\textrm{sys},i}$ are 
the total, statistical and systematic uncertainties
for apparent magnitude, respectively.
The statistical error $\sigma_{\textrm{stat},i}$ in a redshift bin 
consists of several components and is then
\begin{align}
\sigma_{\textrm{stat},i} = \frac{1}{N_{i}}\sqrt{\sigma_{m,i}^2+\sigma_{D}^2+\sigma_{\textrm{lens},i}^2},
\end{align}
where $N_i$ is  the number of SNe for the $i$-th bin,
$\sigma_{m,i}$ is the photometric measurement error per supernova,
$\sigma_{D}$ is the intrinsic dispersion in luminosities of SNe, and  
$\sigma_{\textrm{lens},i}$ is the contribution of gravitational lensing magnifications.
Here, we assume Gaussian uncertainties, and thus  
we add in quadrature these errors in the Fisher matrix above.

In our analysis, we use the specification of WFIRST-AFTA
in Ref.\cite{Spergel:2015sza}, 
which is listed in Table~\ref{tab:SN_spec}.  
For all bins including a near sample, 
we assume the common 
values for the measurement and intrinsic errors as $\sigma_{m,i}=0.08$ and $\sigma_{D} = 0.09$. 
For the contribution of gravitational lensing $\sigma_{\textrm{lens},i}$,
it is modeled as $\sigma_{\textrm{lens},i}=0.07\times \bar{z}_i$. 
Furthermore, we assume the systematic error per bin is given by
$\sigma_{\textrm{sys},i}=0.01\times (1+\bar{z}_i)/1.8$.

In observations of SNe, the absolute magnitude $M$ is treated as a nuisance parameter 
and marginalized away
to reduce the Fisher matrix for the cosmological parameter set.
This procedure means that  the Hubble constant $h$ cannot be determined from SNe, and
it would be redundant to take $\Omega_mh^2$ as a primary parameter. 
In our analysis therefore, we remove $\Omega_mh^2$ from a primary parameter but 
only $\Omega_X$, and the dark energy EoS parameters are treated as primary ones 
for the Fisher matrix of SNe.

%%%%%%%%%%%%%%%%%%%%%%%%%%%%%%%%%%%%%%%%%%%%%%%%%%%%%%%%%%%%%%%

\subsection{Future constraints on dark energy parameters}

\subsubsection{Parametrization I}

\begin{figure}[htbp]
\centering
\includegraphics[bb=10 143 590 660,width=1\linewidth]{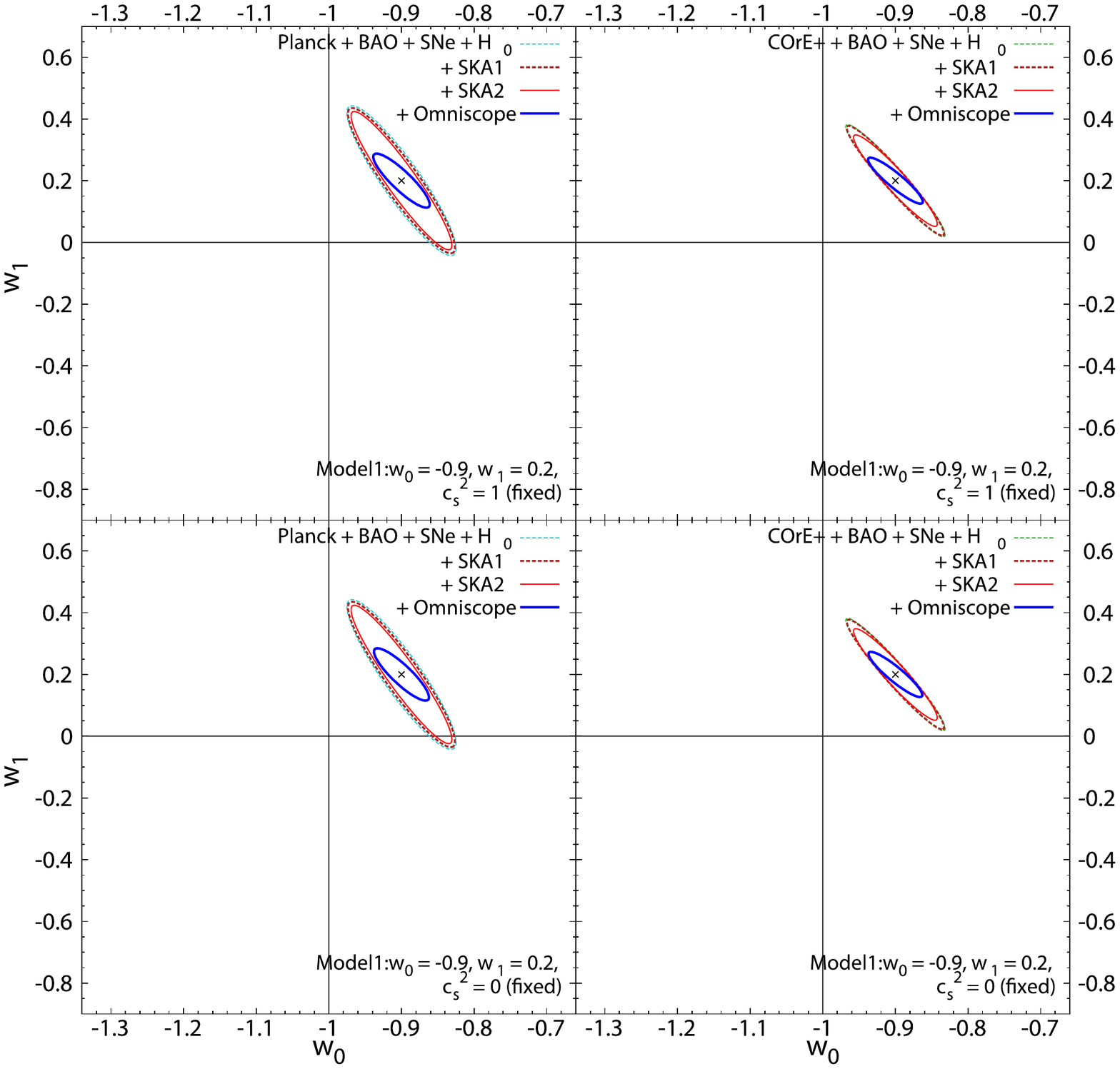}
  \caption{Expected constraints at 95 \% C.L. for the parametrization~I.  
  We assume $w_0 = -0.9$ and $w_1 =0.2$ for the fiducial values.
    Constraints from Planck + BAO + SNe + $H_0$ with SKA or Omniscope (left panel) and
    COrE+ + BAO + SNe + $H_0$ with SKA or Omniscope (right panel) are shown.
The sound speed is taken to be $c_s^2= 1$ (top panels) and $0$ (bottom panels), respectively.}
  \label{fig:CPL_w0w1}
\end{figure}

\begin{figure}[htbp]
\centering
\includegraphics[bb= 20 55 610 760,width=1\linewidth]{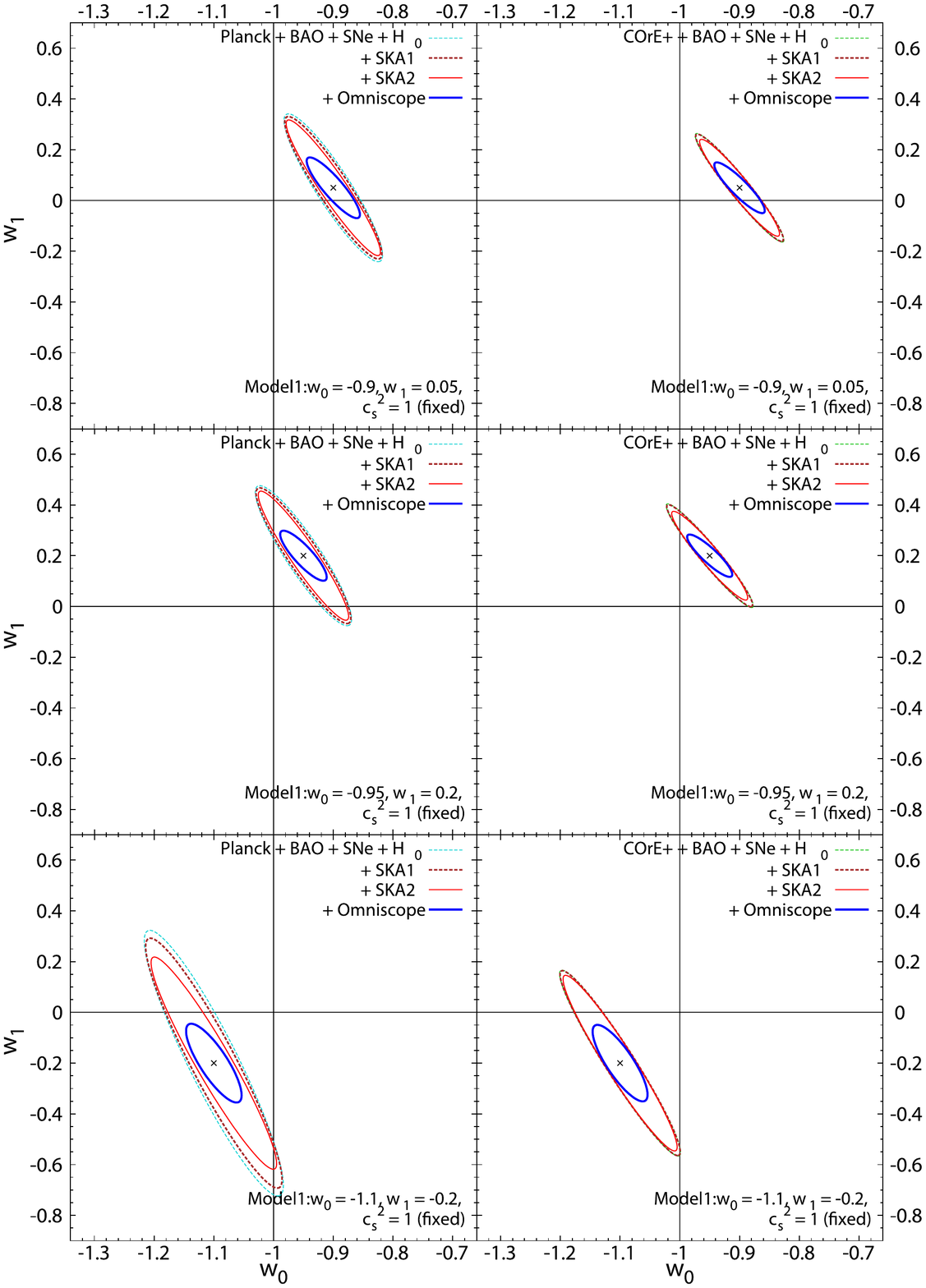}
\vspace{5mm}
  \caption{Expected constraints at 95 \% C.L. for the parametrization~I 
  for the fiducial values of $(w_0, w_1) = (-0.9, 0.05)$ (top),
    $(-0.95, 0.2)$ (middle) and $(-1.1, -0.2)$ (bottom). In this
    figure, we fix the sound speed to be $c_s^2=1$.  }
  \label{fig:CPL_w0w1_2}
\end{figure}

Since there are only two parameters for the EoS in this
parametrization, i.e., $w_0$ and $w_1$, we show expected constraints
on the $w_0$--$w_1$ plane. In Fig.~\ref{fig:CPL_w0w1}, we show
parameter regions allowed at 95 \% C.L. where we used CMB, BAO, SNe and
$H_0$ with SKA1, SKA2 and Omniscope.  For CMB, we performed the analysis
by adopting the specifications of Planck and COrE+.  The fiducial
values for the EoS parameters are assumed to be $w_0= -0.9$ and
$w_1= 0.2$ in Fig.~\ref{fig:CPL_w0w1}, and
$(w_0, w_1) = (-0.9, 0.05), (-0.95, 0.2), (-1.1, -0.2)$ in
Fig.~\ref{fig:CPL_w0w1_2}, respectively.  We also need to specify the
effective sound speed of dark energy $c_s^2$ to follow the evolution
of the linear perturbations\footnote{
The anisotropic stress can also be non-zero if one considers a very
general class of dark energy models~\cite{Ichiki:2007vn}. However,
its effects are not significant either. Hence here we ignore the anisotropic stress of dark energy.
}, which are taken to be $c_s^2 = 1$ or $c_s^2 =0$ in
Fig.~\ref{fig:CPL_w0w1}.  By comparing the constraints for the cases
with $c_s^2=1$ and $0$, one finds that they are almost the same
regardless of the value of $c_s^2$, which indicates that the effects
of varying $c_s^2$ is very weak, and its assumption scarcely affects
the constraints.  This also indicates that the nature of dark energy perturbation 
 cannot be well probed by CMB nor 21 cm fluctuations, at least
for dark energy models with this parametrization, and the constraints
mainly comes from its effects on the background evolution.

In Fig.~\ref{fig:CPL_w0w1_2}, expected constraints are shown for the
fiducial values of $(w_0, w_1) = (-0.9, 0.05)$ (top), $(-0.95, 0.2)$
(middle) and $(-1.1, -0.2)$ (bottom) with the sound speed being fixed
to be $c_s^2=1$.  As seen from Figs.~\ref{fig:CPL_w0w1} and
\ref{fig:CPL_w0w1_2}, when we add information from the 21 cm
fluctuations, the constraints in this parametrization cannot be so
drastically improved for SKA1 and SKA2, but can be done significantly
for Omniscope.  We should stress that the tendency depends on the
parametrization as we discuss in the following sections.  However, for
the parametrization I, although the 21 cm observations look at
 different redshift range from CMB and other observations,  and can probe multiple redshift slices,
we need a better specification at the level of Omniscope to
obtain a severe constraint on dark energy parameters.

\subsubsection{Parametrization II}

\begin{figure}[htbp]
\centering
\includegraphics[bb= -10 40 630 780 ,width=1\linewidth]{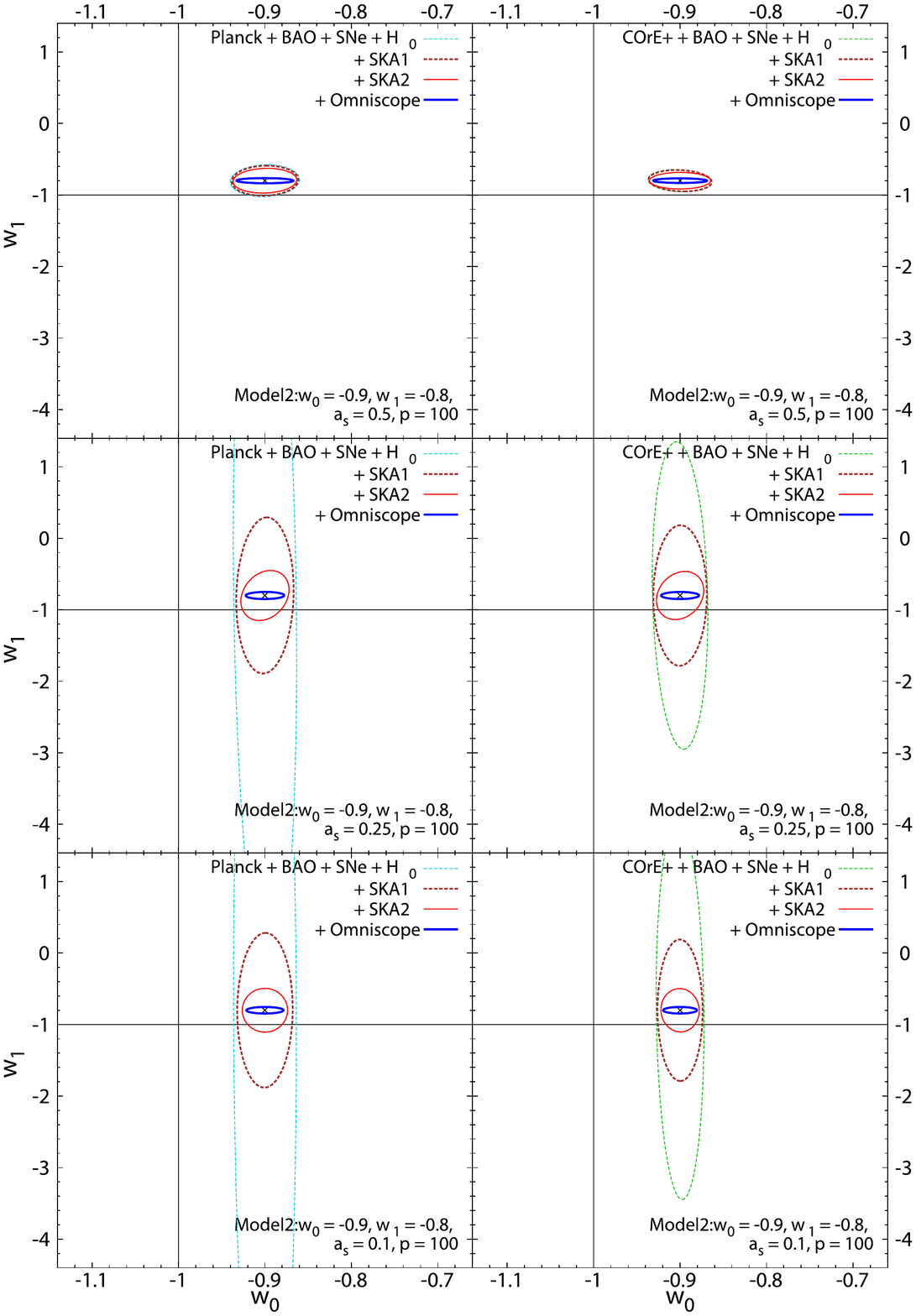}
  \caption{Expected constraints at 95 \% C.L. on the $w_0$--$w_1$
plane for the parametrization~II.  The fiducial values are taken
to be $w_0= -0.9$ and $w_1 = -0.8$.  
Other EoS parameters are
assumed as $a_s =0.5$ (top panels), $0.25$ (middle panels), 
$0.1$ (bottom panels) and $p=100$. All parameters except $w_0$ and $w_1$ are marginalized over. 
In the left and
right panels, Planck and COrE+ are assumed for CMB, respectively.
  }
  \label{fig:HM_w0w1_p100}
\end{figure}

\begin{figure}[htbp]
\centering
    \includegraphics[bb=-10 40 630 780 ,width=1\linewidth]{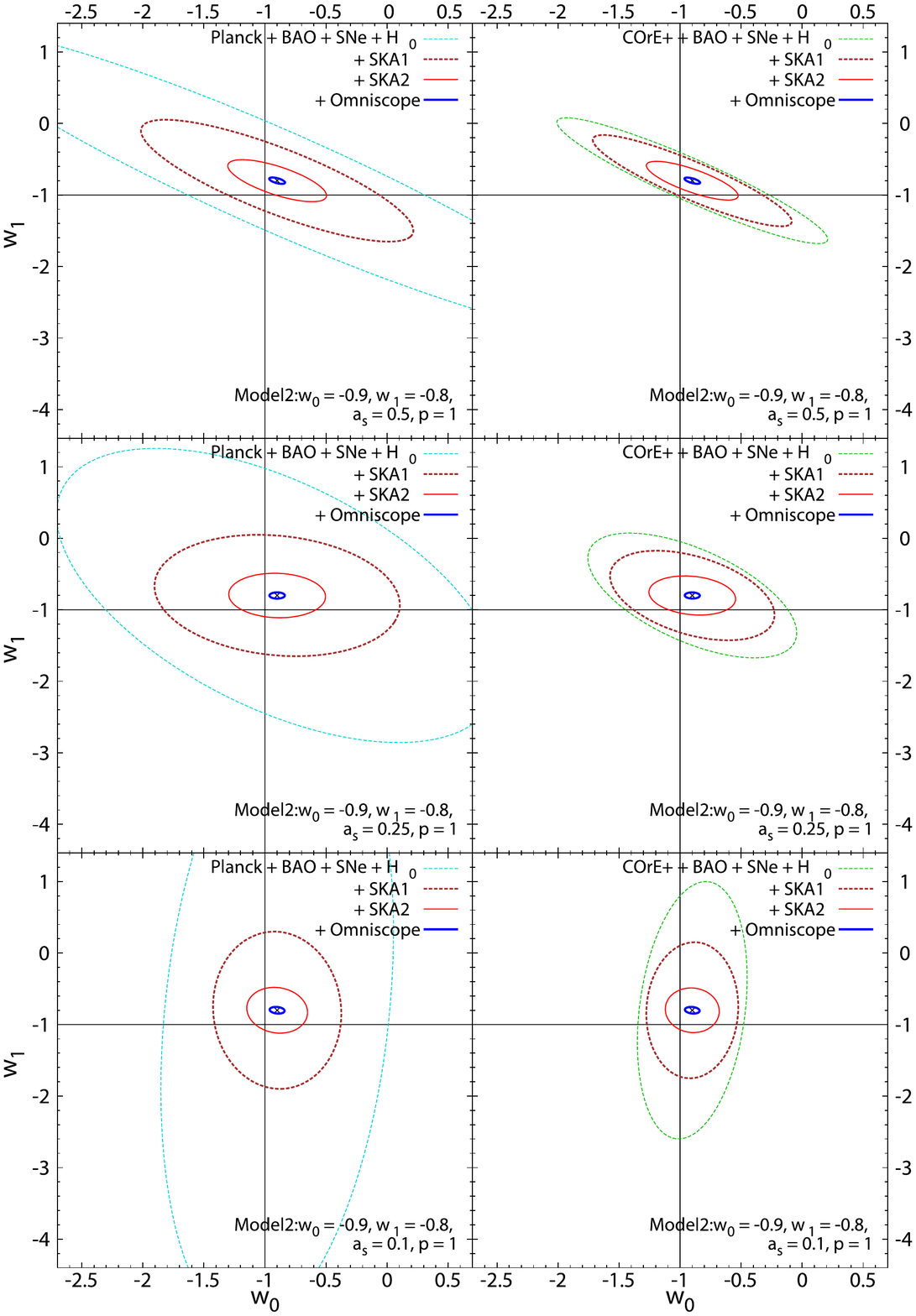}
  \caption{The same as Fig.~\ref{fig:HM_w0w1_p100} but for $p=1$.  }
  \label{fig:HM_w0w1_p1}
\end{figure}

This parametrization includes four parameters to describe the EoS:
$w_0, w_1, a_s$ and $p$.  To present our results, we show the
constraints on 2D planes fixing or marginalizing other parameters.
First we show the expected constraints on the $w_0$--$w_1$ plane in
the cases of $p=100$ in Fig.~\ref{fig:HM_w0w1_p100}, and $p=1$ in
Fig.~\ref{fig:HM_w0w1_p1}.  In each figure, we take several values for
$a_s$ to be $a_s=0.5, 0.25, 0.1$ and assume the fiducial values as
$w_0 = - 0.9$ and $w_1= -0.8$.  As seen from
Figs.~\ref{fig:HM_w0w1_p100} and \ref{fig:HM_w0w1_p1}, the constraints
significantly improve when we add information from the 21 cm
fluctuations, depending on the values of $a_s$ and $p$.

First let us discuss the case of $p=100$. When $p$ is so large, the
EoS abruptly changes at the corresponding scale factor
$a_s$ (or the redshift $z_s$).  In the case of $a_s=0.1$ (or $z_s=9$)
shown in the bottom panel of Fig.~\ref{fig:HM_w0w1_p100}, the EoS 
is almost unchanged after $z=9$. Since BAO and SNe only probe
low redshift $(z \lesssim 2)$, they are sensitive to the EoS at low
redshift.  That is why $w_1$ cannot be determined for the case with
the fiducial value of $a_s=0.1$ with CMB+BAO+SNe+$H_0$.  It should be noted here that
CMB is not so powerful to constrain the EoS compared to BAO and SNe
since integrated quantities such as the angular diameter distance to
last scattering surface are the main probes.  However CMB is very
important to determine the other cosmological parameters which can
break some degeneracies among the parameters including the dark energy
EoS.  Also, for the case with $a_s=0.25$ (or $z_s = 3$), the data set
of CMB+BAO+SNe+$H_0$ cannot well determine $w_1$. In these cases
however, when we add information of 21 cm fluctuations which probes
the redshift $6.8 \le z \le 10$, $w_1$ can be severely
constrained. This shows the power of the 21 cm observations to
investigate dark energy.  In the case of $a_s=0.5$, the EoS changes
from $w_1$ to $w_0$ at $z_s \sim 1$. Therefore, by the observations of
BAO and SNe, we can constrain both of $w_0$ and $w_1$. However even in
this case, when we include SKA2, the constraint on $w_1$ slightly
improves. When Omniscope is included, the constraint becomes more severe.

Next let us consider the case of $p=1$, which is shown in
Fig.~\ref{fig:HM_w0w1_p1}.  Then, the EoS changes slowly from $w_1$ to
$w_0$. Hence even when the transition redshift is higher than $z=2$,
$w_1$ can be constrained severely by low-redshift observations like BAO and SNe.
A slow change of the EoS also indicates that
both of $w_0$ and $w_1$ can be well probed at higher redshift,
irrespective of the transition redshift $z_s$.  Therefore, the 21 cm
observations always improve the constraints in every case for $p=1$ as
shown in Fig.~\ref{fig:HM_w0w1_p1}.

\begin{figure}[htbp]
\centering
\includegraphics[bb=-10 40 630 780,width=1\linewidth]{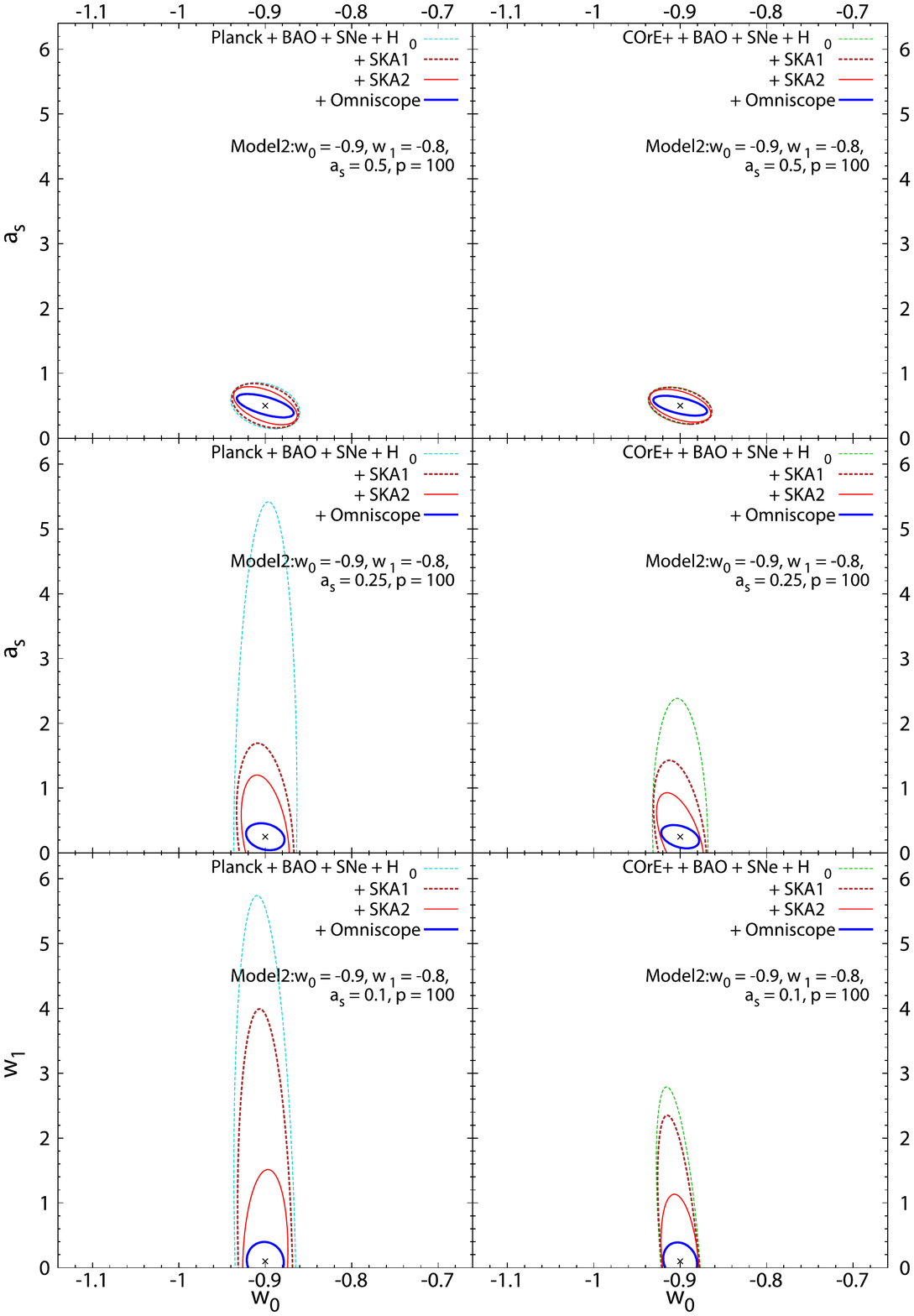}
  \caption{Expected constraints at 95 \% C.L.  on the $w_0$--$a_s$
plane for the parametrization~II. 
For the fiducial values, we
assume $(w_0, w_1) =(-0.9, -0.8)$ and $a_s =0.5$ (top panels),
$0.25$ (middle panels),  $0.1$ (bottom panels) and $p=100$.
All parameters except $w_0$ and $a_s$ are marginalized over.
In deriving these constraints, $w_1$ is marginalized
over.}
\label{fig:HM_w0as_p100}
\end{figure}

\begin{figure}[htbp]
\centering
\includegraphics[bb=-10 40 630 780 ,width=1\linewidth]{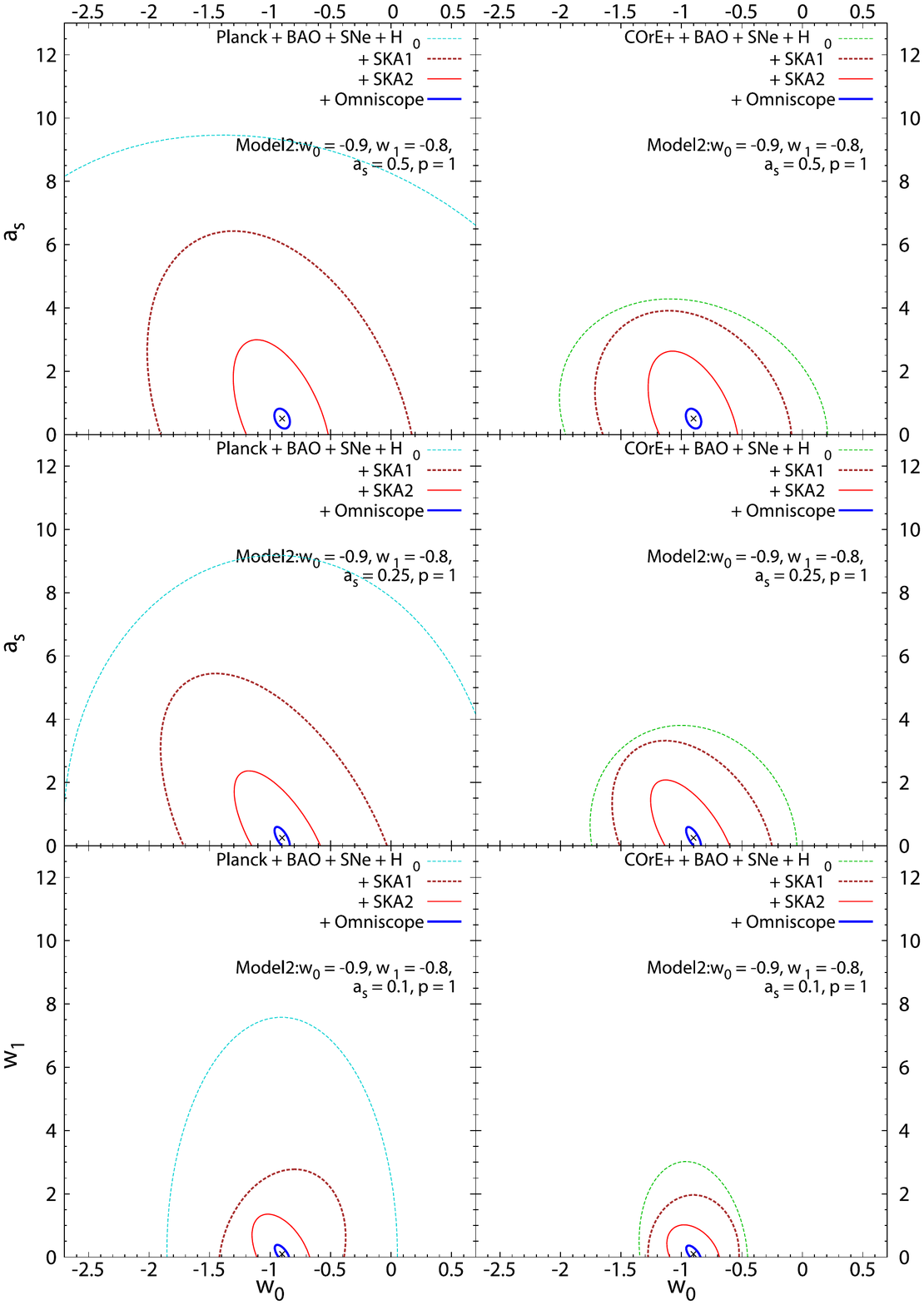}
  \caption{The same as Fig.~\ref{fig:HM_w0as_p100} but for $p=1$.
  }
  \label{fig:HM_w0as_p1}
\end{figure}

Now we discuss the constraints on the $w_0$--$a_s$ plane in the cases
of $p=100$ and $p=1$ in Figs.~\ref{fig:HM_w0as_p100} and \ref{fig:HM_w0as_p1}, respectively. 
For the case of $p=100$, the 21 cm
observations significantly improve the constraints when the transition
redshift is high such as $a_s = 0.1$ and $0.25$ as those in the
$w_0$--$w_1$ plane.  For the case of $p=1$, the tendency is also the
same as those in the $w_0$--$w_1$ plane discussed above.

Regarding the parametrization II, the observations of 21 cm
fluctuations significantly improve the constraints when the transition
redshift is high or the EoS changes slowly.

\subsubsection{Parametrization III}

\begin{figure}[tbp]
\centering
\includegraphics[bb= 10 143 590 660 ,width=1\linewidth]{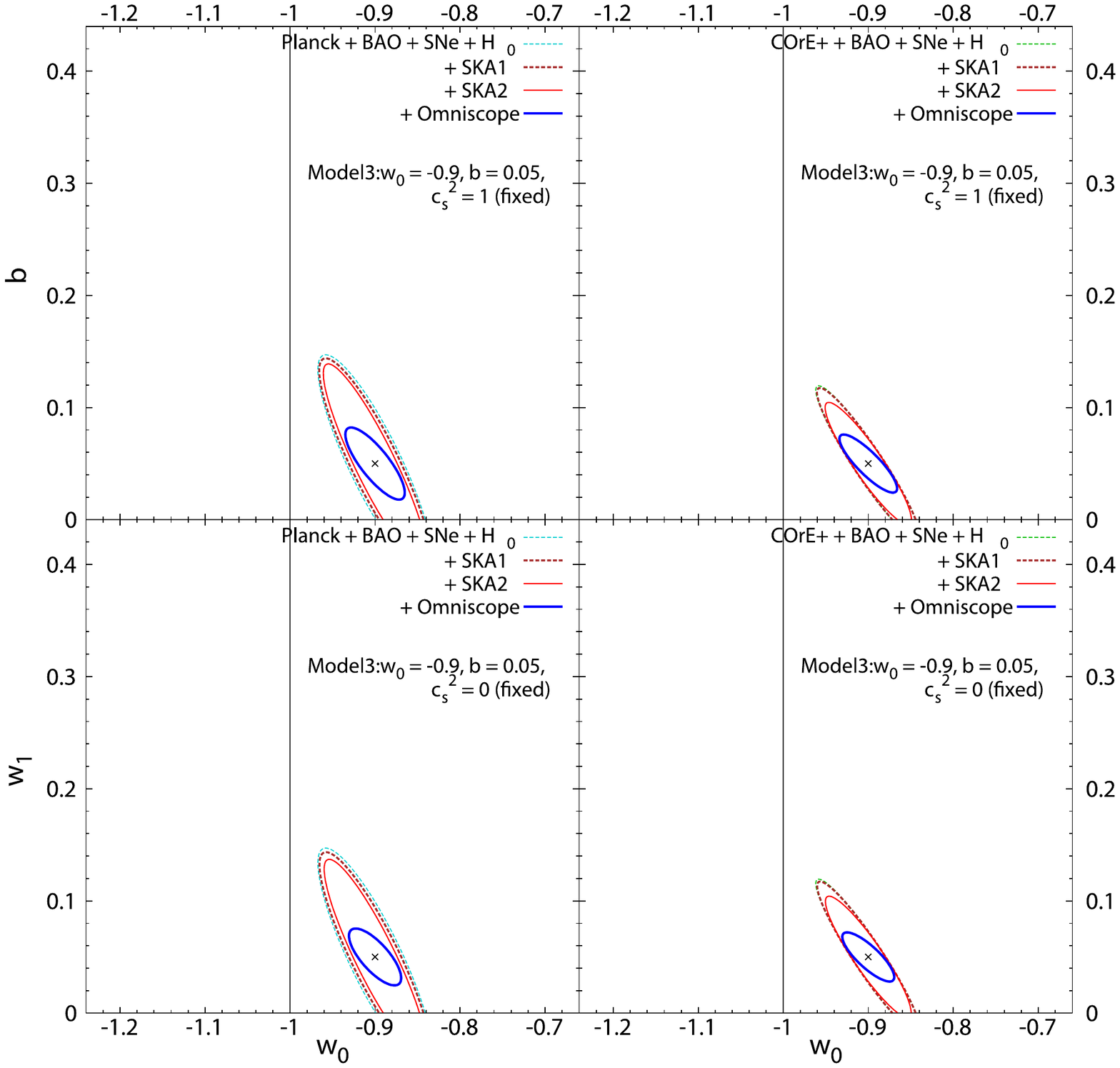}
  \caption{Expected constraints at 95 \% C.L. on the $w_0$--$b$ plane
    for the parametrization~III.  The fiducial values are assumed as
    $w_0 = -0.9$ and $b= 0.05$.  The effective sound speed is fixed to be
    $c_s^2=1$ (top panels) and $c_s^2 =0$ (bottom panels).  }
  \label{fig:Wett_w0b}
\end{figure}

\begin{figure}[htbp]
\centering
\includegraphics[bb= -30 270 650 580 ,width=1\linewidth]{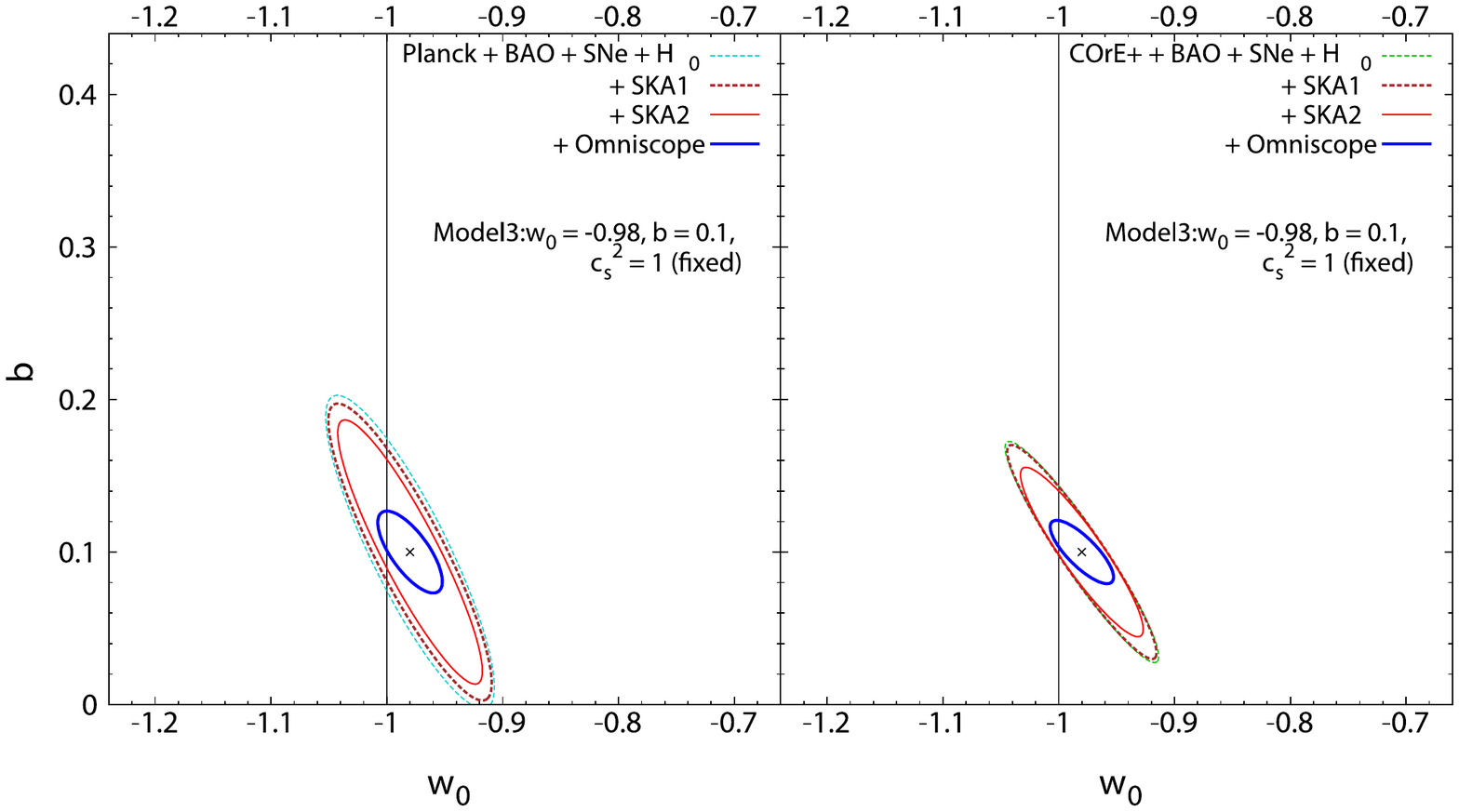}
  \caption{Expected constraints at 95 \% C.L. on the $w_0$--$b$ plane
    for the parametrization III for the fiducial values of
    $(w_0,b)= (-0.98, b= 0.1)$.  In this figure, the effective sound
    speed is fixed to be $c_s^2=1$.  }
  \label{fig:Wett_w0b_2}
\end{figure}

This parametrization includes only two parameters, $w_0$ and $b$.
Hence we show constraints expected from future 21cm and other
observations in the $w_0$--$b$ plane in
Figs.~\ref{fig:Wett_w0b}~and~\ref{fig:Wett_w0b_2}.  For the fiducial
values, we have assumed $w_0 = -0.9$ and $b= 0.05$ in
Fig.~\ref{fig:Wett_w0b}, and $w_0 = -0.98$ and $b= 0.1$ in
Fig.~\ref{fig:Wett_w0b_2}, which are inside the allowed region of the
current constraint presented in the previous section.  The effective
sound speed is taken to be $c_s^2 = 1$ or~$0$.

As discussed in Section~\ref{sec:DE_param}, this parametrization is
suitable to describe the so-called early dark energy model, in which
the effects of the dark energy perturbation can be more significant
than those in other parametrizations.  However, as one can see from
the figure, the constraints for $c_s^2=1$ and $0$ are almost the same,
which indicates that the information on the background evolution
mostly determines the constraints.  Similarly to the parametrization
I, the inclusion of the 21 cm fluctuations do not much
improve the constraints at the level of SKA. However, when Omniscope
is taken into account, the constraints become more severe, which shows
potential of the 21cm fluctuations as a probe of dark energy.

%%%%%%%%%%%%%%%%%%%%%%%%%%%%%%%%%%%%%%%%%%%%%%%%%%
\section{Conclusion and discussion}
\label{sec:conclusion}
%%%%%%%%%%%%%%%%%%%%%%%%%%%%%%%%%%%%%%%%%%%%%%%%%%

%%%%%%%%%%%%%%%%%%%%%%%%%%%%%%%%%
\begin{table}[tbp]
\begin{tabular}{rc|cccc}
\hline \hline
 Model &             & CMB+BAO+SNe+$H_0$                 & + SKA1                    & + SKA2                    & + Omniscope \\ \hline
Parametrization I    & $w_0$ & $ 2.76 \times 10^{  -2} $ & $ 2.70 \times 10^{  -2} $ & $ 2.32 \times 10^{  -2} $ & $ 1.52 \times 10^{  -2} $ \\  
                     & $w_1$ & $ 7.34 \times 10^{  -2} $ & $ 7.17 \times 10^{  -2} $ & $ 5.97 \times 10^{  -2} $ & $ 3.00 \times 10^{  -2} $ \\  \hline
Parametrization II   & $w_0$ & $ 1.11 \times 10^{  -2} $ & $ 1.05 \times 10^{  -2} $ & $ 8.93 \times 10^{  -3} $ & $ 7.90 \times 10^{  -3}  $ \\  
$(p,a_s)=(100,0.1)$  & $w_1$ & $ 1.07                  $ & $ 3.99 \times 10^{  -1} $ & $ 1.22 \times 10^{  -1} $ & $ 1.84 \times 10^{  -2}  $ \\ \hline  
Parametrization II   & $w_0$ & $ 1.29 \times 10^{  -2} $ & $ 1.25 \times 10^{  -2} $ & $ 1.10 \times 10^{  -2} $  & $ 8.85 \times 10^{  -3} $ \\   
$(p,a_s)=(100,0.25)$ & $w_1$ & $ 8.67 \times 10^{  -1} $ & $ 3.96 \times 10^{  -1} $ & $ 1.36 \times 10^{  -1} $  & $ 1.99 \times 10^{  -2} $ \\ \hline  
Parametrization III  & $w_0$ & $ 2.49 \times 10^{  -2} $ & $ 2.43 \times 10^{  -2} $ & $ 2.05 \times 10^{  -2} $  & $ 1.37 \times 10^{  -2} $ \\   
                     & $b$   & $ 2.80 \times 10^{  -2} $ & $ 2.72 \times 10^{  -2} $ & $ 2.20 \times 10^{  -2} $  & $ 1.04 \times 10^{  -2} $ \\  
\hline \hline
\end{tabular}
  \centering 
  \caption{Summary of the expected 1$\sigma$ errors. For CMB, we adopt the specification of COrE+.  The fiducial values for the EoS parameters are assumed to be $(w_0,w_1)=(-0.9,0.2)$ for Parametrization I,  $(w_0,w_1)=(-0.9,-0.8)$ for Parametrization II, $(w_0,b)=(-0.9,0.05)$ for Parametrization III, and the effective sound speed is fixed as $c^2_s=1$.
  }\label{tab:summary}
\end{table}
%%%%%%%%%%%%%%%%%%%%%%%%%%%%%%%%%%

We have investigated expected constraints on dark energy, especially,
the parameters which describe the EoS for several dark energy
parametrizations using future observations of 21 cm line in
combination with CMB, BAO, SNe and $H_0$.  Since the 21 cm line
observations can probe different redshift ranges from CMB, BAO and
SNe, they can give more information combined with those other
observations.  We have assumed the specifications of SKA phase 1, SKA
phase 2 and Omniscope for the future 21 cm observations.  We have
analyzed the three parametrizations for the dark energy equation of
state: the parametrization I, II and III.  Parametrization I given in
Eq.~\eqref{eq:EoS_1} is most commonly used one when the time
dependence of the EoS is investigated.  Parametrization II
(Eq.~\eqref{eq:EoS_2}) can represent the dark energy models which
predict a sudden change in the EoS at some redshift.  Parametrization
III (Eq.~\eqref{eq:EoS_3}) corresponds to the so-called early dark
energy model where its EoS gets close to zero as the redshift
increases. This feature allows the fraction of the energy density of
dark energy can be sizable even at higher redshift.

By assuming these parametrizations, we have studied expected
constraints on the parameters in the EoS with and without the future
21 cm observations to see their 
power to probe the properties of dark
energy.  We performed the Fisher matrix analysis and have investigated 
to what extent the future 21 cm observations can improve the
constraints on the EoS parameters.  We found that a degree of the
improvement depends on the parametrization and the fiducial values of
the EoS.  Table~\ref{tab:summary} summarizes the expected 1$\sigma$
bound on the EoS parameters for each parametrizations.

For the parametrization I, the inclusion of SKA (even phase 2) does
not improve much the constraints. However when Omniscope is combined
with CMB+BAO+SNe+$H_0$, the constraints should be significantly
improved as clearly shown in Fig.~\ref{fig:CPL_w0w1_2}.

For the parametrization II, the observations of SKA can significantly
improve the constraints on the EoS parameters, depending on the
fiducial values. In the parametrization II, the EoS is allowed to
change at some redshift from $w_1$ (the EoS at earlier time) to $w_0$
(the EoS at later time), which introduces the extra two parameters to
describe its evolution: the width and redshift of the transition.
When the width of the transition is very narrow, and the transition
redshift is high, the observations of BAO and SNe cannot probe the EoS
in the earlier time, i.e., $w_1$. However, the 21 cm observations can
probe higher redshift. Therefore the inclusion of the observations of
21 cm fluctuations can significantly improve the constraints, in
particular that for $w_1$.  However, when the transition redshift is
low so that BAO and SNe can probe its change, the constraints from
CMB+BAO+SNe+$H_0$ are already severe enough. Hence not so much
improvement can be made by the 21 cm observations in such a case.  These are
illustrated in Fig.~\ref{fig:HM_w0w1_p100}.  When the width of the
transition is broad, the EoS evolves only gradually, which enables us
to probe the EoS in various redshifts. This means that BAO and SNe can
always be sensitive to both $w_0$ and $w_1$. Therefore they can be
constrained, regardless of the transition redshift without the 21 cm
observations.  However, as shown in Fig.~\ref{fig:HM_w0w1_p1}, the
constraints can be much improved when the observation of 21 cm is
included.  We can also see these tendencies by looking at the
constraints on the $w_0$--$a_s$ plane as shown in
Figs.~\ref{fig:HM_w0as_p100} and \ref{fig:HM_w0as_p1}.  In
constraining dark energy which can be characterized by the
parametrization II, the observations of 21 cm fluctuations are very
powerful.

For the parametrization III, the tendency is similar to that for the
parametrization I. In this parametrization, the 21 cm observations at
the level of SKA is not so powerful to improve the
constraints. However, when we consider Omniscope-type observations,
the determination of the EoS would be much improved, which shows the
potential of the 21 cm fluctuations as a probe of this kind of dark energy.

The power of the 21cm line observations may depend on the actual
properties of dark energy (i.e., the fiducial value of the EoS
parameters and the function form for the time dependence of
EoS). Concretely the 21 cm line observations are powerful especially
when the EoS of dark energy varies at relatively high redshifts, which
are difficult to be probed by other observations.
The known key probes such as galaxy-galaxy lensing and
  redshift-space distortions are definitely important in order to
  constrain dark energy~\cite{Font-Ribera:2013rwa}. Here we would like
  to stress that the 21cm line observations can be complementary to
  those probes because the ranges of the observed redshift are much
  higher.  
In near future, the so-called next generation of the 21
cm observations such as SKA will become available.  Therefore,
investigating dark energy with 21 cm is very timely and should be
pursued further.

%%%%%%%%%%%%%%%%%%%%%%%%%%%%%%%%%%%%%%%%%%%%%%%%%%%%%%%%%%%%%%%%%%%%%%
\section*{Acknowledgments}
%%%%%%%%%%%%%%%%%%%%%%%%%%%%%%%%%%%%%%%%%%%%%%%%%%%%%%%%%%%%%%%%%%%%%%

This work is partially supported by JSPS KAKENHI Grant Number
JP15K05084 (TT), 26247042 (KK), and MEXT KAKENHI Grant Number JP15H05888
(TT), JP15H05889, JP16H0877 (KK).  This work was supported by IBS under
the project code, IBS-R018-D1 (TS).

\pagebreak
\appendix 
\renewcommand{\thetable}{\Alph{section}.\arabic{table}}
\section{Specifications of observations}
\label{sec:spec}

\begin{table}[h]
\begin{tabular}{c|ccccccc}
\hline \hline
          &  $N_{\rm ant}$  &  $A_e~(z=8)$   & $L_{\rm min} $ & $ L_{\rm max}$ & FoV $(z=8)$     &  Obs. time $t_0$ & z \\
          &                         &  $[{\rm m}^2]$   &  $[{\rm m}]$      &   $[{\rm km}]$   &  $[{\rm deg}^2]$ &   [hour]                &  \\ \hline
SKA phase 1 & 911/2         & 443 & 35 & 6 & 13.12 $\times$ 4  & 1000 & 6.8 -- 10 \\
SKA phase 2 & 911$\times$ 4 & 443 & 35 & 6 & 13.12 $\times$ 4  & 1000 & 6.8 -- 10 \\
  Omniscope  & $10^6$ & 1  &  1 & 1 & 2.063 $\times 10^4$ & 1000 & 6.8 -- 10  \\ \hline
\hline
\end{tabular}
\centering 
\caption{
Specifications of 21~cm observations for SKA \cite{ska} and Omniscope \cite{Tegmark:2009kv}.
Note that the effective collecting area $A_e$ and the field of view FoV are proportional to 
$\lambda^2$, where $\lambda$ is the observed wave length.
However, for Omniscope, we assume that $A_e$ and FoV are fixed.
For SKA phase~1 and phase~2, 
multiple fields are assumed to be observed,
and set the number of fields $N_{{\rm field}}$ to be 4 in our analyses.
Then, the effective field of view of SKA phase~1 and phase~2 is
${\rm FoV_{SKA}} = 13.12 \times N_{{\rm field}}$ $[{\rm deg}^2]$.
It should be noted that, although we have assumed the number of fields to be 4, 
our results  are almost unchanged even if we change $N_{\rm field}$ for a fixed total observation time.
(We have checked this by calculating the expected constraint for different numbers of fields $N_{\rm field}$ 
and observation time $t_0$ with the total observation time being fixed (i.e. 4000 hours) and found that 
expected constraints are almost unchanged by the choice of $N_{\rm field}$ and $t_0$.)
}
  \label{tab:SKA_spec}
\end{table}

\bigskip
\begin{table}[h]
\begin{tabular}{c|c|c|c}
\hline \hline 
  Central frequency (GHz)        &     $\theta_{\rm FWHM}$  (arcmin)         &  $\Delta_T$ ($\mu$K arcmin)   &  $\Delta_P$ ($\mu$K arcmin)  \\ 
  \hline
  75 & 14 & 2.7 & 4.7 \\
 105 & 10 & 2.7 & 4.6 \\
 135 & 7.7 & 2.6 & 4.5 \\
 165 & 6.4 & 2.6 & 4.6 \\
  195 & 5.4 & 2.6 & 4.5 \\
   225 & 4.7  & 2.6 & 4.5 \\
\hline \hline
\end{tabular}
  \centering 
  \caption{Specification of COrE+ \cite{core+}. 
  Although there are other channels, we only list those for CMB channels which are used in our analysis.}
  \label{tab:core_spec}
\end{table}

%%%%%%%%%%%%%%%%%%%%%%%%%%%%%%%%%%%%%%%%%%%%%%%%%%%%%
\bigskip
\begin{table}
\begin{tabular}{c|c|c}
\hline \hline 
 Central redshift $z_{i}$ & $\sigma_{d}(z_{i}) \times 10^{2}$  &  $\sigma_{H}(z_{i})\times 10^{2}$ \\ 
\hline
0.15 & 2.78 & 5.34 \\
0.25 & 1.87 & 3.51 \\
0.35 & 1.45 & 2.69 \\
0.45 & 1.19 & 2.20 \\
0.55 & 1.01 & 1.85 \\
0.65 & 0.87 & 1.60 \\
0.75 & 0.77 & 1.41 \\
0.85 & 0.76 & 1.35 \\
0.95 & 0.88 & 1.42 \\
1.05 & 0.91 & 1.41 \\
1.15 & 0.91 & 1.38 \\
1.25 & 0.91 & 1.36 \\
1.35 & 1.00 & 1.46 \\
1.45 & 1.17 & 1.66 \\
1.55 & 1.50 & 2.04 \\
1.65 & 2.36 & 3.15 \\
1.75 & 3.62 & 4.87 \\
1.85 & 4.79 & 6.55 \\
\hline \hline
\end{tabular}
  \centering 
  \caption{Specification of DESI. We reproduce Table~5 in \cite{Font-Ribera:2013rwa} only including the relevant quantities 
  for our analysis. Note that $\sigma_d (z_i)$ and $\sigma_H (z_i)$ are the errors of $\ln (d_A (z_i))$ and $\ln (H(z_i))$, respectively.}
  \label{tab:DESI_spec}
\end{table}
%%%%%%%%%%%%%%%%%%%%%%%%%%%%%%%%%%%%%%%%%%%%%%%%%%%%%%%

%%%%%%%%%%%%%%%%%%%%%%%%%%%%%%%%%
\bigskip
\begin{table}[tbp]
\begin{tabular}{c|cccccccccccccc}
\hline \hline
$z_{\textrm{max}}$      & 1.7   & 1.6 & 1.5 & 1.4 & 1.3 & 1.2 & 1.1 & 1.0  \\
$z_{\textrm{min}} $     & 1.6   & 1.5 & 1.4 & 1.3 & 1.2 & 1.1 & 1.0 & 0.9  \\ 
$N_{i}                  $ & 136 & 136 & 136 & 136 & 136 & 136 & 136 & 136  \\ \hline
$z_{\textrm{max}}$      & 0.9   & 0.8 & 0.7 & 0.6 & 0.5 & 0.4 & 0.3 & 0.2 & 0.1 \\
$z_{\textrm{min}} $     & 0.8   & 0.7 & 0.6 & 0.5 & 0.4 & 0.3 & 0.2 & 0.1 & 0.03 \\ 
$N_{i}                  $ & 136 & 136 & 136 & 326 & 223 & 402 & 208 & 69  & 800 \\
\hline \hline
\end{tabular}
  \centering 
  \caption{
The number of SNe for each bin assumed in our analysis, which is adopted in \cite{Spergel:2015sza} for WFIRST-AFTA.
In the same manner as presented in the report \cite{Spergel:2015sza},
we include a ``near sample'' of 800 SNe 
and the last bin corresponds to it.
The low redshift supernovae are 
observed by ground based experiments,
and we assume that 
their statistical, and systematic errors are 
the same as those in the far sample. 
For the statistical errors,
we assume $ \sigma_{m,i}=0.08$, $\sigma_{D}=0.09$ and  $\sigma_{\textrm{lens},i}=0.07\times \bar{z}_i$. 
}\label{tab:SN_spec}
\end{table}
%%%%%%%%%%%%%%%%%%%%%%%%%%%%%%%%%%

\pagebreak

\end{document}